\documentclass[prl,twocolumn,aps,showpacs,amsmath,amssymb,superscriptaddress,floatfix,longbibliography]{revtex4-1}

\usepackage{color}
\usepackage[dvipsnames]{xcolor}
\usepackage{bm}
\usepackage{graphicx}
\usepackage{braket}
\usepackage[belowskip=-9pt]{caption}
\usepackage{subcaption}
\usepackage{ragged2e}
\DeclareCaptionJustification{justified}{\justifying}
\graphicspath{{Figs/}}
\newcommand{\bS}{{\bf S}}

\usepackage[pdftex]{hyperref}
\usepackage{cleveref}
\hypersetup{colorlinks = true, urlcolor = blue, linkcolor = blue, citecolor = blue}

\begin{document}
\title{Magnon Bose-Einstein Condensation and Superconductivity in a Frustrated Kondo Lattice}
\author{Pavel A. Volkov}
\affiliation{Department of Physics and Astronomy, Center for Materials Theory, Rutgers University, Piscataway, NJ 08854 USA}
\author{Snir Gazit}
\affiliation{Racah Institute of Physics and The Fritz Haber Research Center for Molecular Dynamics, The Hebrew University, Jerusalem 91904, Israel}
\author{J. H. Pixley}
\affiliation{Department of Physics and Astronomy, Center for Materials Theory, Rutgers University, Piscataway, NJ 08854 USA}
\date{\today}

\begin{abstract}
Motivated by recent experiments on magnetically frustrated heavy fermion metals, we theoretically study the phase diagram of the Kondo lattice model with a nonmagnetic valence bond solid ground state on a ladder. A similar physical setting may be naturally  occurring in YbAl$_3$C$_3$, CeAgBi$_2$, and TmB$_4$ compounds. In the insulating limit, the application of a magnetic field drives a quantum phase transition to an easy-plane antiferromagnet, which is described by a Bose-Einstein condensation of magnons. Using a combination of field theoretical techniques and density matrix renormalization group calculations we demonstrate that in one dimension this transition is stable in the presence of a metallic Fermi sea and its universality class in the local magnetic response is unaffected by the itinerant gapless fermions. Moreover, we find that fluctuations about the valence bond solid ground state can mediate an attractive interaction that drives unconventional superconducting correlations. We discuss the extensions of our findings to higher dimensions and argue that, depending on the filling of conduction electrons, the magnon Bose-Einstein condensation transition can remain stable in a metal also in dimensions two and three.
\end{abstract}

\maketitle
Correlated metals with closely competing quantum ground states provide an important platform for studying a range of fascinating phenomena such as strange metallicity~\cite{lohnheysen.2007,sachdev.2011}, unconventional superconductivity \cite{scalapino.2012}, and fractionalized excitations~\cite{coleman.2001,sachdev.2008}. An important example thereof is the Kondo lattice model realized in heavy fermion materials \cite{wirth.2016}, where the competition between magnetic order and screening of local moments induces a
non-Fermi liquid in the vicinity of a quantum critical point \cite{coleman.2001,SiSteglichReview}. More recently, it has been pointed out \cite{si.2010,coleman.2010} that Kondo systems with frustrated local moments open a largely unexplored
avenue of quantum criticality beyond the Ginzburg-Landau-Wilson paradigm, where screening competes with a {\it quantum disordered} spin state, such as a spin liquid \cite{zhou.2017} or a static-crystalline pattern of local singlets, i.e. a valence bond solid (VBS) \cite{Zapf-2014}. 
The recent discoveries of heavy fermion metals with local moments residing on geometrically frustrated lattices; e.g. the Shastry-Sutherland lattice in Yb$_2$Pt$_2$Pb, Ce$_2$Pt$_2$Pb, and Ce$_2$Ge$_2$Mg (Ref.~\cite{KimAronson3}), a distorted triangular lattice in YbAl$_3$C$_3$ (Ref.~\cite{Hara-2012}), and a distorted Kagome lattice in CeRhSn~\cite{Tokiwa-2015} and CePd$_{1-x}$Ni$_x$Al~\cite{LohneysenCe}, provide an excellent platform to study the interplay of magnetic frustration and metallicity.

From a theoretical perspective, it is necessary to establish what properties of frustrated magnetism,
including  quantum critical phenomena, are stable in the presence of a metallic band. This question is delicate, as both the magnetic
fluctuations and the electronic excitations
are gapless at the critical point. In this work, we consider one of the best understood transitions in insulating quantum magnets, the so-called magnon Bose-Einstein condensation (BEC)~\cite{Giamarchi-2008,Zapf-2014}. 
It occurs in insulating frustrated 
VBS 
magnets, such as TlCuCl$_3$ (Refs.~\cite{Oosawa-1999,Nikuni-2000}) and SrCu$_2$(BO$_3$)$_2$ (Ref.~\cite{Miyahara-2003}),
where the application of a magnetic field 
destroys the local singlets to produce
an ordered state of spin triplets. The measured critical properties of this transition agree well with the BEC universality class. There are several material candidates to host the magnon BEC phenomena in a metal: YbAl$_3$C$_3$~\cite{Hara-2012}, CeAgBi$_2$~\cite{Thomas-2016}, and TmB$_4$~\cite{Shin-2017}. Each of these compounds either have supporting experimental evidence of a valence bond solid ground state in the absence of a field or magnetization plateaus that acquire a finite slope that we expect is due to the Zeeman coupling to the conduction band. Moreover, in YbAl$_3$C$_3$ the 
field tuned
transition to a magnetically ordered phase~\cite{Khalyavin-2013} is accompanied by a logarithmic behavior of specific heat that has been interpreted as a signature of a non-Fermi liquid \cite{Hara-2012}.

\begin{figure}[h!]
	\begin{subfigure}[b]{0.4\textwidth}
		\phantomcaption
		\label{fig:ladder}
	\end{subfigure}
	\begin{subfigure}[b]{0.4\textwidth}
		\phantomcaption
		\label{fig:bands}
	\end{subfigure}
	\begin{subfigure}[b]{0.4\textwidth}
		\phantomcaption
		\label{fig:phase_diag}
	\end{subfigure}
	\includegraphics[width=\linewidth]{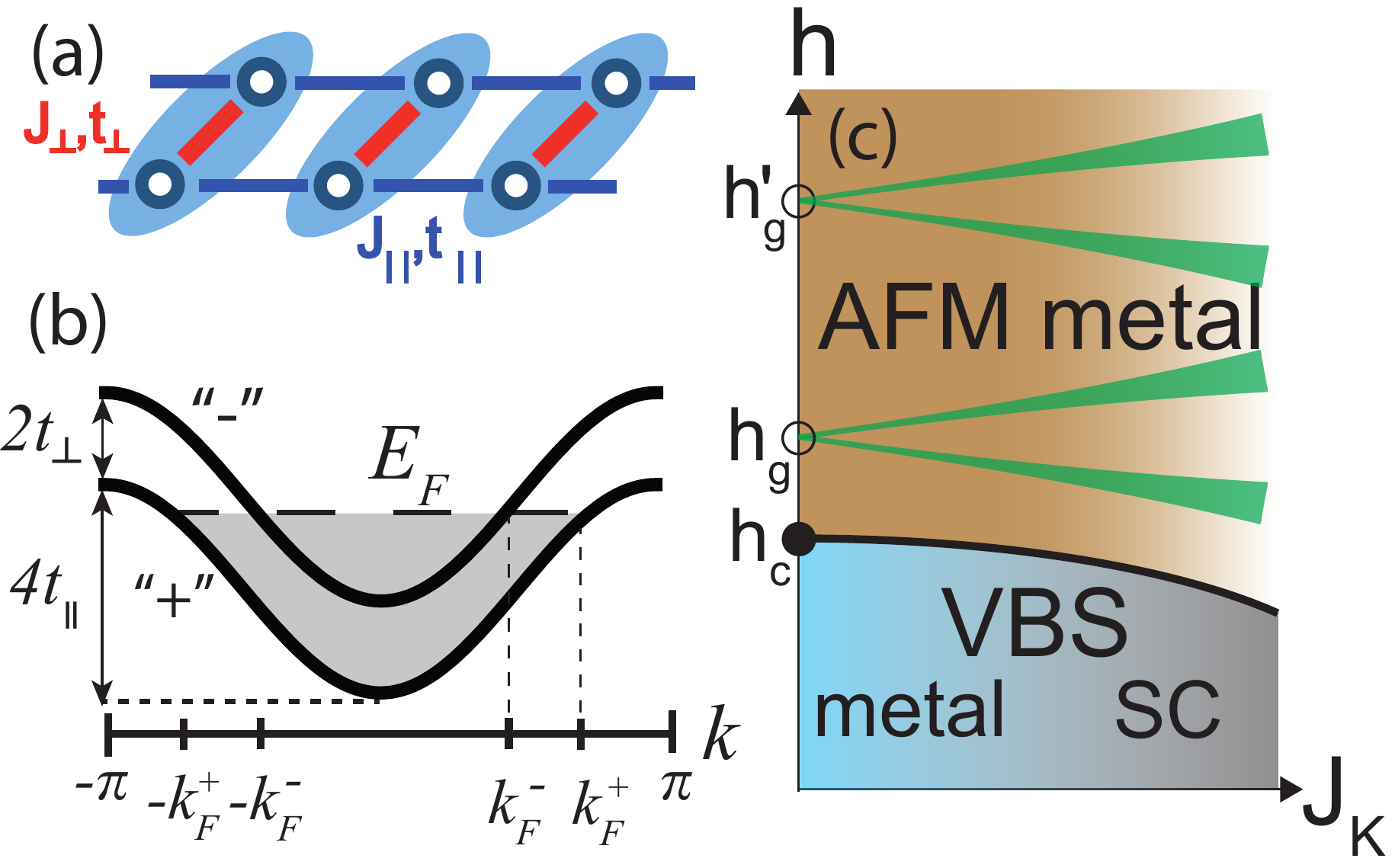}
	\caption{
		(a) Schematic depiction of the model studied. Each site hosts a localized spin and a conduction electron site coupled to their nearest neighbors. Shaded ovals represent the localized spin singlets in the VBS phase.  (b) The band structure of conduction electrons for $t_\perp<2t_\parallel$. Depending on the filling, one or two bands can cross the Fermi level. (c) Schematic phase diagram of the model. The VBS-AFM transition occurs at $h_c$ 
		and is of the BEC universality class. In the AFM phase partial gaps open near $h_g$ and $h_g^{'}$ determined by the filling  at nonzero $J_K$ (shaded green areas). For two bands crossing the Fermi level superconductivity (SC) emerges for sufficiently large $J_K$.
	}
\end{figure}

Currently, it is unknown if the magnon BEC transition can take place in a metal. To address this question, we study a  lattice model~\cite{giamarchi.1999} exhibiting a magnon BEC transition to an easy-plane (XY) antiferromagnetic (AFM) phase in its insulating limit and solve it in the presence of a metallic conduction band, see figs.~\ref{fig:ladder} and~\ref{fig:bands}. Using a combination of a low-energy field theoretical analysis and density matrix renormalization group (DMRG) calculations we present a comprehensive solution to the problem in one dimension (1D) and demonstrate  that
the magnon BEC transition is stable in a metal. 

Our main results are summarized in \cref{fig:phase_diag}. We find that the VBS state and the BEC transition survive in the presence of a metallic conduction band. For the case with two partially filled bands we find that superconducting correlations induced by spin fluctuations develop. 
In the AFM phase, the Kondo interaction induces
partially gapped 
regimes for
certain values of the magnetic field. 
Finally, we show that for 
a single partially filled band,
the stability of the magnon BEC transition carries over to two- (2D) and three-dimensional (3D) generalizations of the model. This allows us to provide a clear-cut theoretical explanation for the observed non-Fermi liquid scaling in YbAl$_3$C$_3$.

{\bf The model:}
We consider the Kondo-Heisenberg model on the two-leg ladder (see \cref{fig:ladder}), governed by the Hamiltonian $H=H_c+H_f+H_K$. $H_c$ contains the single-particle dispersion of the conduction band, $H_f$ describes the interacting spin-1/2 local moments, and $H_K$ corresponds to the Kondo coupling between the two. For $H_f$ we consider the Heisenberg model on a ladder~\cite{giamarchi.1999},
\begin{equation}
\begin{small}
\begin{gathered}
H_f = J_\perp\sum_r {\bf S}_{r,1}\cdot{\bf S}_{r,2} +
J_\parallel\sum_{r,\alpha} {\bf S}_{r,\alpha}\cdot {\bf S}_{r+1,\alpha}
- h g_f \sum_{r,\alpha} \bS^z_{r,\alpha},
\end{gathered}
\end{small}
\label{eq:hf0}
\end{equation}
where ${\bf S}_{r,\alpha}$ denotes a vector of spin-1/2 operators at site $r$, chain $\alpha=1,2$. For $J_\perp\neq 0$ \cite{white1994,shelton1996} the ground state at $h=0$ is adiabatically connected to a direct product of singlets on each rung and thus serves as a minimal model of the VBS state. This state does not break any symmetries of the model in \eqref{eq:hf0}; however, at $h>h_c$ a transition to a state with a quasi-long range ordered AFM state occurs \cite{giamarchi.1999} (see also below). Thus, the magnetic field $h$ allows to tune the ground state of the local moments from a quantum disordered (VBS) state to a more conventional symmetry-breaking one (AFM), in full analogy to the frustration parameter proposed in Refs. \cite{si.2010,coleman.2010}. In what follows, we set the $g$-factor for the local moments to $g_f=1$. For analytic calculations we concentrate on the regime $J_\perp\gg J_\parallel$ in a finite $h\sim J_\perp$. 
In this regime, the two low-energy states on each rung are the singlet and the lowest lying triplet, whereas the other two triplet states are separated by an energy gap of size $\sim J_\perp$. The low-energy sector can be exactly mapped onto either hard-core bosons (denoted as $a_r$ for an annihilation operator) or spinless fermions (denoted as $f_r$ and referred to as spinons)~\cite{giamarchi.1999,giamarchi.2003}. In either representation an occupied site is equivalent to a rung in the triplet state, while an empty site is a singlet along the rung.

The conduction electron Hamiltonian reads $
H_c = \sum_{k,p=\pm} E_p(k)\psi^\dagger_{k,p}\psi_{k,p}
$
where the dispersion is given by $E_{\pm}(k)= -2 t_\parallel \cos{k}  \mp t_{\perp} - \mu$, for a chemical potential $\mu$, 
the lattice constant is set to unity, and $\psi_{k,\pm} = (\psi_{k,1} \pm \psi_{k,2})/\sqrt{2}$ are two-component spinors in the bonding/antibonding
basis. The resulting band structure is presented in \cref{fig:bands}. For the low-energy properties of the system, it is important whether the Fermi energy crosses both bands [as in \cref{fig:bands}] or only one, which we will refer to as two- and one-band cases, respectively. As the localized spins are usually due to $f$-electrons with a large total angular momentum~\cite{carlin.1986} as compared to the conduction electrons (often from $s$ or $d$ states) we have omitted the Zeeman term in $H_c$. Below we will argue that relaxing this approximation does not fundamentally change any of our main conclusions.

Finally, the conduction electrons interact with the localized spins via an antiferromagnetic Kondo coupling $H_K = J_K \sum_{r, \alpha}{\bf S}_{r,\alpha}\cdot {\bf s}_{r,\alpha}$ where ${\bf s}_{r,\alpha}=\psi^{\dag}_{r,\alpha} (\bm{\sigma}/2)\psi_{r,\alpha}$ and $J_K>0$. 
To make headway analytically, we project the ${\bf S}_{r,\alpha}$ operators in $H_K$ onto the low-energy sector of Eq. \ref{eq:hf0} and obtain in the hard-core boson representation
\begin{equation}
\begin{gathered}
H_K \approx \frac{J_K}{4} \sum_r(a_r^\dagger a_r) (\psi^\dagger_{r,+}\sigma_z\psi_{r,+}+\psi^\dagger_{r,-}\sigma_z\psi_{r,-})
\\
-\frac{J_K}{2\sqrt{2}} \sum_r
[
a_r(
\psi^\dagger_{r,+}\sigma^+\psi_{r,-}+\psi^\dagger_{r,-}\sigma^+\psi_{r,+}
)+h.c.
],
\end{gathered}
\label{eqn:Hk}
\end{equation}
One sees that the spin-flip term acts only {\em between} the two fermion bands. As is shown below, this has important consequences, namely stabilizing the BEC transition against Kondo screening.

{\bf Numerical methods:} For the numerical solution of the model, we use the DMRG algorithm as implemented in the ITensor package \cite{ITensor}, targeting the low-lying states and their physical properties. The presence of gapless and near-critical modes requires a careful numerical analysis to avoid a bias toward low-entangled states. To that end, we have monitored the convergence of DMRG results as a function of bond dimension, keeping up to 9830 states for system sizes up to $L=76$ rungs (see SI Appendix, section 6 for details).

{\bf Magnon BEC transition:} We first consider the one-band case for fields below the BEC transition $h<h_{c}$. As the second band is gapped, one can integrate the fermions out (See section 2 of SI Appendix). Ignoring the hard-core constraint, we find the leading terms (in $J_K$) to renormalize the bosonic spectrum. In particular, the critical field is reduced
\begin{equation}
\begin{gathered}
h_c(J_K)  \approx h_c(0) -\frac{J_K^2}{8 t_\perp}
\int_{0}^{k_F}\frac{dk}{\pi}
\frac{1}{1+2\frac{t_\parallel}{t_\perp}\cos(k)}
\\
-\frac{J_K^2}{64 \pi^2}\int_{|k|>k_F} \int_{-k_F}^{k_F} dk dk' \frac{1}{J_\parallel\sin^2\frac{k-k'}{2}-t_\parallel[\cos (k) - \cos (k')]}
,
\end{gathered}
\label{eq:deltahc1}
\end{equation}
where the expression is for the case of the bottom band being partially filled and $J_K^2/t_\perp\ll J_\parallel$ is assumed (see SI Appendix for details). The top line of \eqref{eq:deltahc1} derives from the transverse part of the Kondo coupling (\eqref{eqn:Hk}, bottom line), whereas the bottom line arises from the longitudinal part (\eqref{eqn:Hk}, top line).

Up to second order in $J_K$ we find that the total magnetization $M^z = M_f^z + M_c^z \equiv \sum_{r,\alpha} \langle S^z_{r,\alpha} \rangle + \langle s^z_{r,\alpha} \rangle$ vanishes for fields $h < h_c(J_K)$. However, $M_f^z$ and $M_c^z$ themselves do not vanish. In particular, this results in a Zeeman splitting for the conduction electrons. The nonzero values of $M_f^z$ and $M_c^z$ appear in second order in $J_K$ and do not exhibit any singularities (see SI Appendix for details). This effect remains qualitatively similar in the two-band case.

To determine the critical properties at the transition $h_c(J_K)$ we use the fermionic representation of $H_f$. In this representation the BEC transition corresponds to a Lifshitz transition, where the chemical potential touches the bottom of the spinon band. At the critical point, we take the low-energy continuum limit (with $x$ denoting position) 
to find that the lowest-order interaction term in the gradient expansion $(f^\dagger f)^2$ has to vanish  due to the Pauli principle, whereas all of the higher-order terms [such as $(f^\dagger \partial_x f)^2$] are irrelevant at the QCP due to the additional derivatives~\cite{sachdev.1999}. Thus, interactions that arise due to
integrating out the conduction electrons (see SI Appendix for details) do not affect the transition. The critical low-energy field theory is then that of free fermions with a Lagrangian density
\begin{equation}
\mathcal{L}_{\mathrm{crit}}=f^\dagger(x,\tau) [\partial_\tau - a \partial_x^2 -\mu_f(h)]f(x,\tau),
\label{eq:crit}
\end{equation}
where $f(x,\tau)$ is the spinon field in the continuum, $\tau$ is imaginary time, and $\mu_f(h) =h- h_c(J_K)$.

\begin{figure}[h!]
	\centering
	\begin{subfigure}[b]{0.4\textwidth}
		\phantomcaption
		\label{fig:chi_vs_h}
	\end{subfigure}
	\begin{subfigure}[b]{0.4\textwidth}
		\phantomcaption
		\label{fig:DeltaLsqr_h}
	\end{subfigure}
	\begin{subfigure}[b]{0.4\textwidth}
		\phantomcaption
		\label{fig:hc_vs_JK}
	\end{subfigure}
	\begin{subfigure}[b]{0.4\textwidth}
		\phantomcaption
		\label{fig:chi_vs_L}
	\end{subfigure}
	\includegraphics[width=\columnwidth]{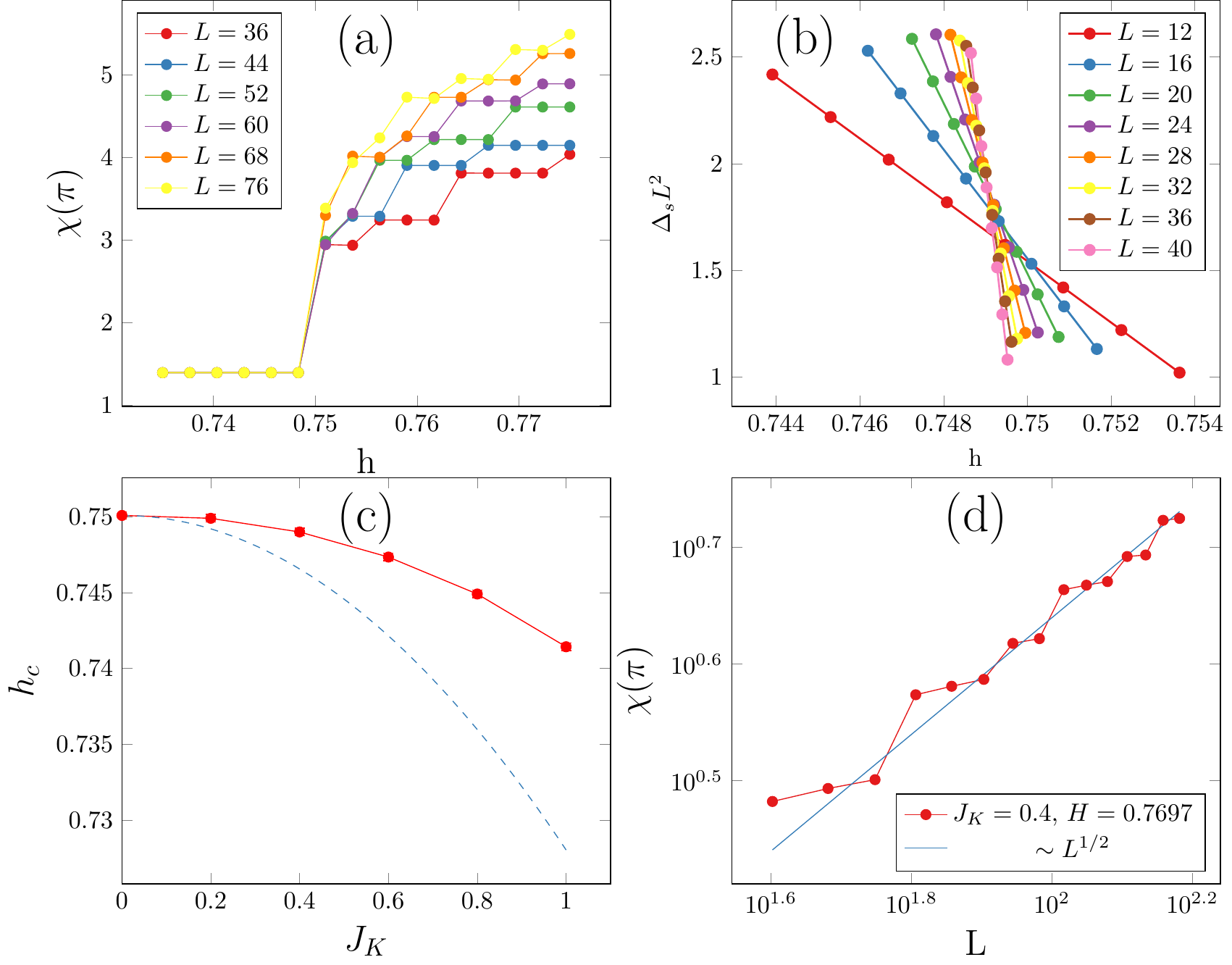}
	\caption{Results of DMRG calculations for the 1-band case, for $J_\perp=1,J_\parallel=0.3,t_\perp=1.5,t_\parallel=1.0$, $J_K=0.4$, and $1/8$ filling. (a) $\chi(\pi)$ as a function of magnetic field $h$ for various system sizes. (b) Curve crossing analysis of the universal amplitude $\Delta_s L^2$ as a function of $h$. (c) Values of the critical field extracted from the location of the curve crossing in (b) for various $J_k$. The dashed line depicts the perturbative result of \cref{eq:deltahc1} computed neglecting the hard-core constraint (d) $\chi(\pi)$ versus system size on a log-log scale for a field close to $h_{c}(J_K=0.4)$. For comparison, solid line depicts the scaling prediction of \cref{eq:crit}.
		\label{fig:dmrg_one_band}}
\end{figure}

There are several predictions that directly follow from the above discussion that we now confirm with DMRG. First, we examine whether the magnon BEC transition survives at finite Kondo coupling. To that end, we set $J_k=0.4$ and compute the transverse spin susceptibility at the ordering wavevector $\chi(\pi) \equiv \sum_{r} e^{i\pi (r-L/2)}\langle (S_{r,1}^+-S_{r,2}^+)(S_{L/2,1}^--S_{L/2,2}^-)\rangle$, see \cref{fig:chi_vs_h}. Indeed, for $h<h_c(J_K)$, the localized spins are short-range correlated and hence $\chi(\pi)$ saturates to a value independent of $L$. On the other hand, for $h>h_c(J_K)$, quasi-long-range order leads 
to a scaling $\chi(\pi)\sim L^\beta$, with $\beta={1-1/(2K)}$ for $h>h_c(J_K)$ \cite{chitra.1997,giamarchi.1999} and a Luttinger parameter $K$ in the AFM phase.

To precisely locate the critical field $h_c$ and determine the critical exponents associated with the transition, we study the finite-size scaling of the spin gap $\Delta_s$ close to criticality. The observable $\Delta_s L^z$ has a vanishing scaling dimension, and consistent with the magnon BEC we assume a dynamical exponent $z=2$. Near criticality this implies that $\Delta_s L^z \sim R[(h-h_c)L^{1/\nu}]$ for an arbitary scaling function $R$ and
that $\Delta_s L^z$ is $L$ independent at $h=h_c$, i.e. $\Delta_s L^z$ versus $h$ curves for an increasing set of systems-sizes precisely cross at a single point identified with the critical field $h_c$.
This relation is tested in \cref{fig:DeltaLsqr_h}, where
we observe a clear crossing in  $\Delta_s L^2$ versus $h$ for  increasing $L$ thus providing an accurate estimate of $h_c$.
We use this unbiased method to accurately determine $h_c$ as a function $J_k$, see \cref{fig:hc_vs_JK}, and find a quadratic decrease  $h_c(J_K=0)-h_c(J_K)\sim J_K^2$, in good qualitative agreement with the free-boson result of \cref{eq:deltahc1}. 

To estimate the correlation length exponent $\nu$, we study the finite-size scaling of $R$-curves at criticality (see section 6 of SI Appendix for details). We find $\nu=0.49(1)$, in compliance with the predicted magnon-BEC universality class value, $\nu_{\mathrm{th}}=1/2$ \cite{sachdev.2008}. Lastly, close to criticality, the free fermion result implies a Luttinger parameter $K =1$ \cite{giamarchi.1999}, which gives $\beta = 1/2$. Again, we find excellent agreement with the DMRG calculation at a field close to $h_c(J_K)$, as shown in \cref{fig:chi_vs_L}, thus further confirming the universality class of the magnon BEC transition in a metal.

We now argue that the BEC transition also remains stable in the two-band case. While the boson self-energy in this regime could have 
divergences at $q=\pm (k_F^+ \pm k_F^-)$, generally, these momenta are not equal to the ordering wavevector $Q=\pi$. Assuming $J_K\ll J_\parallel$, the bosons in the vicinity of $q=\pi$ are then expected to be described by the same theory Eq. \ref{eq:crit} in the vicinity of $h_c(J_K)$ as in the one band case. Hence, the universality class of the transition is unchanged, and corrections to Eq. \ref{eq:deltahc1} are subleading at weak coupling.

{\bf Superconductivity in the 2-band case:} We now consider the properties of conduction electrons for the two-band case (\cref{fig:bands}). We determine the emergent phases using bosonization~\cite{giamarchi.2003}, with the low-energy excitations of the two bands being described with the real bosonic fields $\varphi_{\sigma}^\pm(x)$ and $\varphi_{\rho}^\pm(x)$ for the spin and charge sectors, respectively [each field also has a canonically conjugate one - $\theta^\pm_{\sigma,\rho}(x)$]. In the absence of a Kondo coupling, the low-energy excitations of each of these fields are described as a Luttinger liquid with $K=1, u=v_F\equiv 2 t_\parallel \sin k_F$.

Ignoring for the moment the aforementioned Zeeman splitting, integrating out the gapped hard-core bosons leads to two interaction terms
\begin{equation}
H_{I,\pm}=\frac{J_K^2}{8\varepsilon_{k_F^+\pm k_F^-}\pi^2\tilde\alpha^2}\int dx
\cos \sqrt{2}(\theta^+_\rho - \theta^-_\rho)
\cos \sqrt{2} (\varphi_\sigma^+ \mp \varphi_\sigma^-),
\label{eq:2band:hcos}
\end{equation}
where the index $+(-)$ refers to individual bands (see \cref{fig:bands}), $1/\tilde\alpha$ is a high-energy cut-off, and $\varepsilon_{k_F^+\pm k_F^-}$ is the boson dispersion at $q=k_F^+\pm k_F^-$. Semiclassically, the terms in Eq. \ref{eq:2band:hcos} create a pinning potential for the fields making their excitations gapped. To assess their possible impact in the quantum regime, we performed one-loop renormalization group (RG) analysis. The corresponding equations have been derived in Ref.~\onlinecite{penc.1990}; we take 
the perturbatively generated interactions, including \cref{eq:2band:hcos},
as the initial conditions and solve the equations numerically (see section 3 of SI Appendix for details).
We find that terms proportional to $\cos(2\sqrt{2}\varphi^\pm_\sigma)$ are generated and flow to strong coupling together with the ones in Eq. \ref{eq:2band:hcos}, allowing for a semiclassical analysis. The remaining terms lead to a renormalization of the Luttinger parameters. Minimizing the action including the cosine terms, we find that, out of four gapless fermion modes, only the excitations of the total charge mode $\varphi_\rho^- +\varphi_\rho^+$ remain gapless, corresponding to a state with power-law correlations of the superconducting (SC) order parameter $O_{\mathrm{SC}}^\pm(x)=\psi_{\uparrow,\pm}\psi_{\downarrow,\pm}(x)$  and the conduction electron density at $2k_F^\pm$.
Furthermore, the equilibrium values of the gapped fields are such that the superconducting correlations are sign-changing between the bands, i.e. there is a $\pi$ phase shift between the SC order parameters of the $+$ and $-$ bands, which is an analogue of $d$-wave pairing on the ladder. Additionally, the dominant velocity renormalization is such that the SC correlations are stronger, i.e. decay slower, then the $2k_F^\pm$ density ones. These results resemble the case of two-leg Hubbard ladders \cite{balents.1996} that have $d$-wave superconducting and charge density wave correlations. Similar results have also been obtained for the Kondo-Heisenberg model away from the VBS limit (with $J_\perp=J_\parallel$) in zero magnetic field \cite{xavier2008}.

Let us now discuss the interplay of the above effects and the Zeeman splitting $E_Z$ due to a finite $J_K$. The Zeeman splitting $E_Z$ can thwart superconductivity \cite{chandrasekar.1962,clogston.1962} unless the SC gap $\Delta_{\mathrm{SC}}$ is sufficiently larger then $E_Z$ (for an alternative discussion see SI Appendix). As the interactions are marginal, the gap is expected to be exponentially small $\Delta_{\mathrm{SC}} \sim v_F/\tilde\alpha \exp[-1/g]$, where $g\sim v_F J_K^2/(\tilde\alpha\varepsilon_{k_F^\pm})$, whereas $E_Z \propto J_K^2$. It follows that for infinitesimal $J_K$, $E_Z$ dominates
while at larger $J_K$, $\Delta_{\mathrm{SC}}$
takes over, which represents a ``Doniach-like'' competition between Zeeman splitting and superconductivity. This is similar to the competition between the Kondo coupling and the RKKY interaction.

We note that the superconducting pairing due to VBS fluctuations differs from the conventional scenario of a spin density wave QCP \cite{scalapino.2012}, as well as the one in the 1D Hubbard \cite{balents.1996,dolfi2015} or $t-J$ model. Unlike the spin-density wave QCP \cite{lohnheysen.2007,scalapino.2012}, the magnon BEC critical mode has $z=2$ even without an interaction with fermions, while the 1D Hubbard model does not posses the VBS subsystem. In the case of the $t-J$ model, the binding of holes is achieved due to the energy cost $J_\perp$ of breaking a singlet bond \cite{dagotto1992,dagotto.1996}. The difference from the results for the $t-J$ model  is that the VBS fluctuations are not completely local, especially for $h$ close to $h_c$ and no requirements on the magnitude of $t_\perp$ results \cite{dagotto1992}.

\begin{figure}[t!]
	\centering
	\begin{subfigure}[b]{0.4\textwidth}
		\phantomcaption
		\label{fig:DeltaS_vs_JK}
	\end{subfigure}
	\begin{subfigure}[b]{0.4\textwidth}
		\phantomcaption
		\label{fig:chiD_chiS}
	\end{subfigure}
	\begin{subfigure}[b]{0.4\textwidth}
		\phantomcaption
		\label{fig:nk}
	\end{subfigure}
	\begin{subfigure}[b]{0.4\textwidth}
		\phantomcaption
		\label{fig:EE}
	\end{subfigure}
	\includegraphics[width=\linewidth]{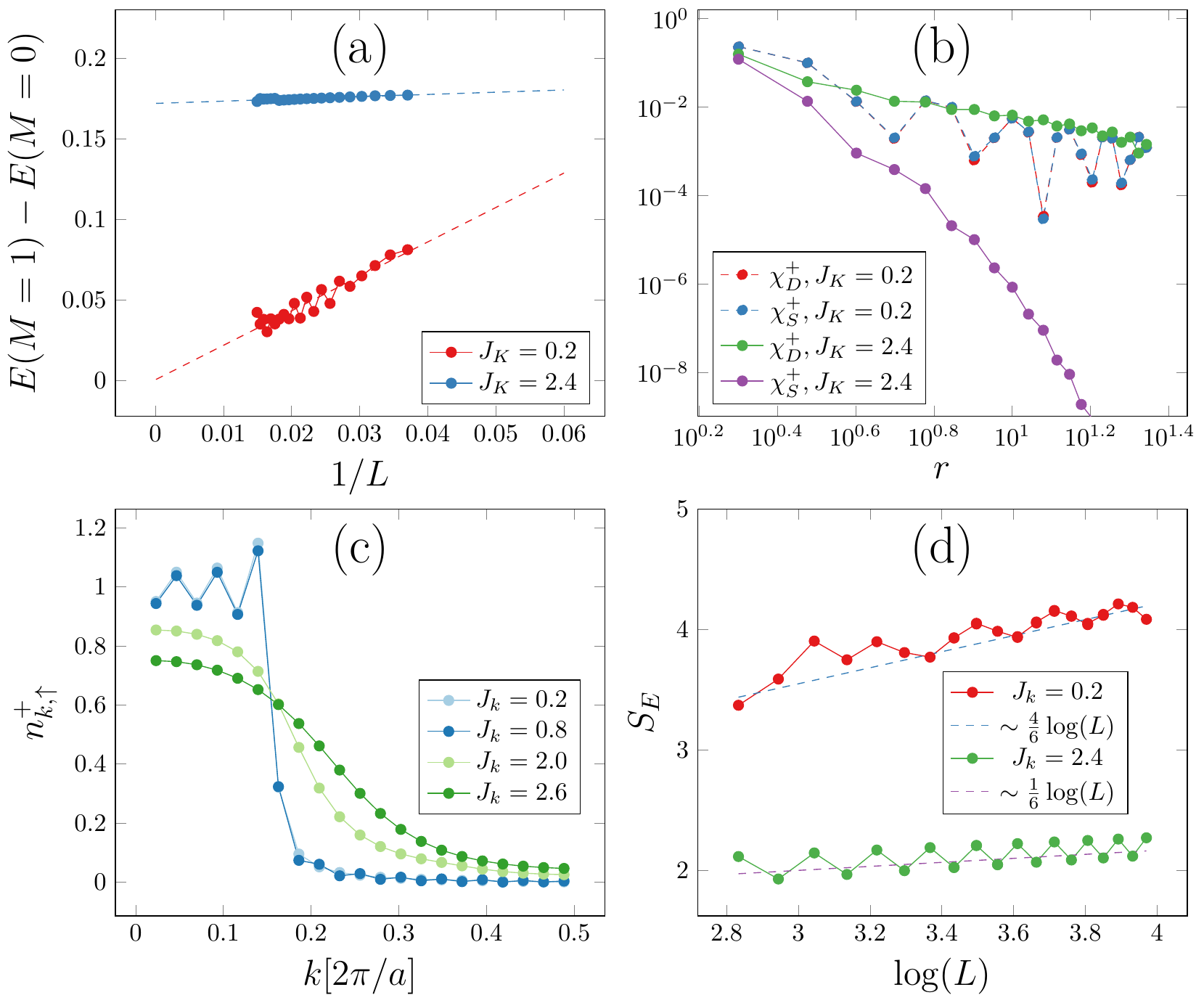}
	\caption{DMRG results for the 2-band case, for $J_\perp=1,$ $J_\parallel=0.3,t_\perp=0.1,t_\parallel=1.0,h=0.5$ and $1/4$ filling, for various $J_K$ in the metallic VBS and superconducting spin gapped phase (a) The spin gap, extracted from a finite size scaling of the energy level spectroscopy, $E(M=1)-E(M=0)$.  Curves are smoothed using a moving average. Dashed line is a fit to a linear function in $1/L$. (b) SDW ($\chi_S$) and SC ($\chi_D$) correlation functions on a log-log scale, for $L=44$.  (c) The spin-resolved momentum occupation number $n^+_{k,\uparrow}$ of the bonding band, for $L=44$. (d) The extraction of the central charge from the scaling of the bipartite von-Neumann entanglement entropy versus the logarithm of system size $\log(L)$. }
	\label{fig:dmrg2}
\end{figure}

The expectations above are confirmed by the DMRG results in \cref{fig:dmrg2}. First, to estimate the SC gap, we obtain the spin gap $\Delta_s$, in \cref{fig:DeltaS_vs_JK} with a finite size scaling analysis of the energy difference $E(M^z=1,L)-E(M^z=0,L)$, between the ground states in the $M^z=1$ and $M^z=0$ sectors (see SI Appendix for details). Indeed, we find a nearly vanishing spin gap $\Delta_s\approx0$ for weak Kondo coupling, $J_k=0.2$, and a finite gap, $\Delta_s\approx 0.17$, at larger Kondo coupling, $J_k=2.4$. To further characterize the above phases, in \cref{fig:chiD_chiS}, we investigate the $c$ electrons intra-band spin-density wave (SDW), $O^\pm_{\mathrm{SDW}}=(\psi^\dagger_{r,1,\downarrow}\pm\psi^\dagger_{r,2,\downarrow})(\psi_{r,1,\uparrow}\pm\psi_{r,2,\uparrow})$, and SC (defined above), $O^\pm_{\mathrm{SC}}$, order parameters, through their respective correlation functions $\chi^+_{\mathrm{D/S}}(r)=\left\langle (O^+_{\mathrm{SC/SDW}})^\dagger(L/2)
O^+_{\mathrm{SC/SDW}}(r)\right\rangle$. At $J_k=0.2$, we find that both order parameters fall off like a power-law, akin to the decoupled free-electron limit. By contrast, in the spin gapped phase, $J_k=2.4$, the SDW correlation decays exponentially while the SC correlation remains quasi-long range.

Next, in \cref{fig:nk}, we examine the spin resolved momentum distribution of the bonding band, $n^{+}_{k,\uparrow} = \langle \psi^{\dag}_{k,\uparrow} \psi_{k,\uparrow} \rangle$. The distribution evolves from a sharp Fermi edge, at small $J_k$, to an incoherent distribution, characteristic of a Luttinger liquid, upon approach to the spin gapped phase. Notably, the Fermi-wavevector is unchanged throughout this transition, unlike the usual Kondo lattice model \cite{schiller.2011,khait.2018}, due the number of spins per unit cell being even (i.e. 2)~\cite{Oshikawa.2000,coleman.2010}.
Finally, using the scaling of the bipartite entanglement entropy  $S_E \sim c/6 \log L$ for a conformal field theory with central charge $c$~\cite{Calabrese-2009}, we find four gapless channels ($c=4$)  in the VBS metal. In the superconducting spin-gapped phase, there is only a single gapless channel ($c=1$) corresponding to the  total charge mode. These results are consistent with the expectations from weak-coupling RG.

{\bf Interactions in the AFM phase:}
For $h > h_c(J_K)$ there is a finite density of spinons in the fermionic representation of the Hamiltonian \cref{eq:hf0}. The low-energy excitations around the spinon Fermi points at $k_F^f$ are of the Luttinger liquid type~\cite{giamarchi.1999} 
that can be described using bosonization. 
The Luttinger parameters $K$ and $u$ are known functions of $h$ \cite{giamarchi.1999}. One can now rewrite the low-energy part of the Kondo coupling in terms of the bosonic fields resulting in three contributions $H_K\cong  H_V+H_Z+ H_I$, where we use $\varphi(x)$ for the spinon fields. First, a velocity renormalization appears due to $H_V =\frac{J_K}{2\sqrt{2}\pi^2} \int dx\partial_x\varphi \partial_x \varphi_\sigma$. $H_I$, on the other hand, describes the interaction between fermionic spinons and the Fermi sea
\begin{equation}
\begin{gathered}
H_{I}
=
\frac{J_K}{4(\pi\tilde{\alpha})^2} \int dx
\cos(2k_F^f x-2\varphi  )
\\
[\cos(2 k_F x - \sqrt{2} (\varphi_\rho+\varphi_\sigma))
-\cos(2 k_F x - \sqrt{2} (\varphi_\rho-\varphi_\sigma))],
\end{gathered}
\label{eq:1band:hcos}
\end{equation}
where $k_F^f$ is the Fermi wavector of the spinons. As $\varphi,\varphi_{\rho,\sigma}$ are slowly varying functions of $x$ the integral averages to zero except for two special cases $k_F^f =k_F,\;\pi-k_F$. If that is so, however, this term is relevant throughout the AFM phase\footnote{The scaling dimension of this term is $1-K + O({J_K/v_F})$ with $K<1$ throughout the AFM phase~\cite{giamarchi.1999}.} and it pins the values of two bosonic fields resulting in a spectral gap of the order $\frac{v_F}{\tilde\alpha} \left(\frac{J_K}{v_F/\tilde\alpha}\right)^{\frac{1}{1-K}}$, similar to the opening of the spin-density wave gap in itinerant magnets.

The presence of $H_Z=-\frac{J_K}{2\sqrt{2}\pi} \int dx M^z_f \partial_x \varphi_\sigma$ term leads additionally to a Zeeman splitting for the conduction electrons. For $h>h_c$, $M^z_f$ is nonzero even in the absence of Kondo coupling and thus this term is linear in $J_K$ and is parametrically larger then both the AFM gap and the perturbatively induced Zeeman splitting $\sim J_K^2$ discussed for $h<h_c$. Thus, Zeeman splitting $\sim J_K$ is the dominant effect of the Kondo coupling and we need to reconsider the effect of the interaction Eq. \ref{eq:1band:hcos} starting with the Zeeman-split Fermi points. The condition for Eq. \ref{eq:1band:hcos} not to be averaged to zero is then $k_F^f =k_F\pm \frac{J_K(\tilde{m}+1/2)}{4 v_F},\;\pi-k_F\pm \frac{J_K(\tilde{m}+1/2)}{4 v_F}$. In each case, one of the  three gapless modes is gapped out. At finite $J_K$ one expects each of these special values broaden into an interval, as is shown by the shaded green regions in Fig. \ref{fig:phase_diag}.

\begin{figure}[h]
	\includegraphics[width=0.48\textwidth]{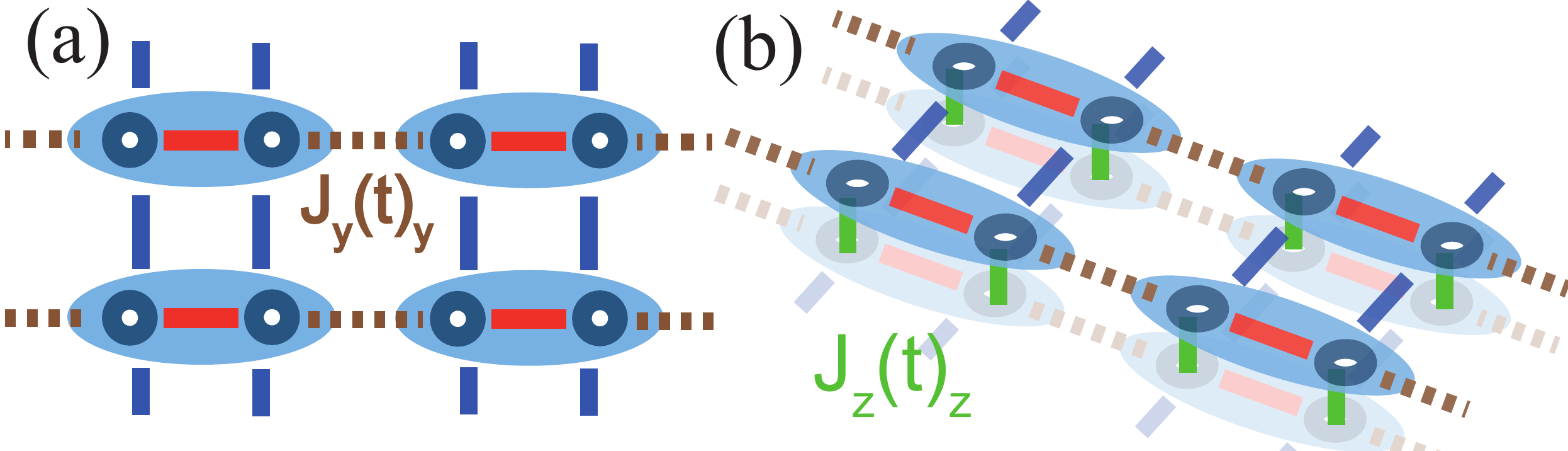}
	\caption{Schematic depiction of the (a) 2D  and (b) 3D extensions of the model in Fig. 1 (a).}
	\label{fig:2d3d}
\end{figure}

{\bf Conduction electron g-factor:}
While throughout we have neglected the $g$-factor, $g_c$ of the conduction electrons, it is a reasonable assumption for a range of fields $g_c h \ll J_K$ thus we consider our results to carry over to the $g_c\neq 0$ case, albeit in finite fields we expect the superconducting correlations to be additionally suppressed.

{\bf Extensions to 2D and 3D:} We now show that the magnon BEC transition is also stable at finite $J_K$ in the one-band case for 2D and 3D extensions of the model considered here. The two-dimensional extension of the model in \cref{fig:ladder} consists of the ladders arranged in a columnar pattern with a weak interladder coupling, and for 3D we stack the resulting layers on top of one another (see Fig. \ref{fig:2d3d}). While the conduction electrons form two bands as before, now the Kondo interaction projected onto the low-energy states possesses an additional intra-band term in addition to \cref{eqn:Hk}
\begin{equation}
H_{K,\pm}^{\mathrm{intra}}=
\frac{J_K}{4\sqrt{2}}
\sum_{{\bf k},{\bf q}}
f({\bf k},{\bf q})
a({\bf q})
\psi^\dagger_{\pm,{\bf k+q}}\sigma^+\psi_{\pm,{\bf k}}
+h.c.
,
\label{eq:2d3d}
\end{equation}
where for $q_y$ close to $\pi$, $f({\bf k},{\bf q})\sim f({\bf k},\pi)(q_y-\pi)$ (see SI Appendix for details). Assuming that the BEC occurs for bosons with $q=Q_0$ [e.g. $Q_0=(\pi,\pi,\pi)$ 
in 3D], only fermions around the Fermi surface points connected by $Q_0$ ('hot spots') are coupled to the critical mode. To assess the influence of the interactions, we calculate their scaling dimension for two cases with the Fermi velocities at the 'hot spots' being antiparallel or not. We define the scaling dimension of energy and the momentum parallel to the Fermi velocities to be $1$: $[\varepsilon] = [k_\parallel] =1$, while the remaining momenta have scaling dimension $2$ to keep $z=2$ intact \cite{sachdev.1999,yamamoto.2010}. Under these assumptions, $H_{K,\pm}^{\mathrm{intra}}$ (Eq.~\ref{eq:2d3d}) as well as the first term of $H_K$ (Eq.~\ref{eqn:Hk}) are irrelevant in $d>1$. For the case of antiparallel [noncollinear] Fermi velocities at the 'hot spots', we find that the former has a scaling dimension $(1-d)/2$ [$-d/2$] and the latter $1-d$ [$-d$]. This provides a strong indication that the BEC transition should retain its universality class in the one-band case.

The above result implies, that the quantum critical behavior is governed by the same theory as in the undoped case, i.e. that of a dilute $z=2$ Bose gas. Interestingly, this prediction may be verified in YbAl$_3$C$_3$, where a VBS ground state of the Yb moments \cite{ochiai2007} is formed on a deformed triangular lattice~\cite{matsumura2008}, while the conductivity suggests metallic behavior \cite{ochiai2007}. Application of a magnetic field results in a quantum phase transition \cite{Hara.2012}, with the specific heat having been found to exhibit a $C_V/T\sim\log(1/T)$ behavior close to the QCP. Initially, this behavior has been attributed to a possible non-Fermi liquid state formed due to Kondo coupling \cite{Hara.2012}. However, our results suggest that the field-induced transition should not be affected by Kondo coupling; indeed for a $z=2$, $d=2$ BEC a $\log(1/T)$ divergence is expected \cite{Fisher.1988,Millis.1993}. Thus, our results allow one to interpret the observed anomalies as a signature of the stability of the magnon BEC transition in metallic systems.

{\bf Discussion and conclusions:}

In this work we have studied quantum critical properties and phases of a one-dimensional frustrated Kondo lattice with a nonmagnetic valence bond solid state in magnetic field. We have shown that the field-induced magnon BEC transition that occurs in the insulating limit is stable in the presence of a metallic conduction band and retains its universality class, with the critical value of the field being lowered. We have demonstrated that VBS fluctuations lead to unconventional superconductivity in the case of two partially filled bands. Finally, we have shown that the stability of the magnon BEC transition extends to higher-dimensional versions of the model. Our results allows us to draw conclusions regarding field-induced transitions in heavy-fermion materials with spins residing on frustrated lattices. In particular, our results have lead us to a clear interpretation of the observed criticality in YbAl$_3$C$_3$ and anchor future studies of the phase diagrams and quantum criticality of frustrated Kondo lattices. It will be also interesting to extend our theory to VBS states that may break a crystalline symmetry as well as field-tuned VBS to AFM transitions that occur at fractional magnetization plateaus~\cite{Miyahara-2003}.

\begin{acknowledgments}

The authors acknowledge useful discussions with D. T. Adroja, M. Aronson, A. Auerbach,  P. Coleman, P. Goswami,  G. Kotliar, E. J. K\"onig, S. Parameswaran, S. Shastry, and Q. Si. 
S.G. and J.H.P. were supported by Grant No. 2018058 from the United States-Israel Binational Science Foundation (BSF), Jerusalem, Israel. S.G. and J.H.P. performed part of this work at the Aspen Center for Physics, which is supported by NSF Grant No. PHY-1607611. P.A.V. acknowledges the support by the Rutgers University Center for Materials Theory Postdoctoral fellowship. DMRG calculations were performed using the ITensor package \cite{ITensor}. Numerical computations were carried out at Intel Labs Academic Compute Environment.

\end{acknowledgments}

\newpage

\begin{widetext}

		\section{Supplemental Material: Magnon Bose-Einstein Condensation and Superconductivity in a Frustrated Kondo Lattice }

	\setcounter{page}{1}
	\renewcommand{\theequation}{S\arabic{equation}}
	\setcounter{equation}{0}
	\renewcommand{\thefigure}{S\arabic{figure}}
	\setcounter{figure}{0}
	
	
In the Supplementary Material we present the details of analytic calculations as well as additional numerical details supporting the results presented in the main text. For the case of a single band crossing the Fermi energy we show how the expressions in the main text are derived; in the AFM phase we present the  bosonized Hamiltonian for the localized spins and discuss the details of the calculation  when the spinon-electron interaction is relevant; for the case when two bands cross the Fermi energy the details of the renormalization group (RG) calculations are given. Finally, we discuss the effects of a commensurate filling of the conduction bands, which was not discussed in the main text.

\tableofcontents

\section{Notations and mapping of spins to spinons/hard-core bosons}

The Kondo-Heisenberg ladder model that we have focused our study on is given by
\begin{equation}
\begin{gathered}
H=H_f+H_c+H_K;
\\
H_f = J_\perp\sum_r {\bf S}_{r,1}{\bf S}_{r,2} + J_\parallel\sum_{r,\alpha} {\bf S}_{r,\alpha} {\bf S}_{r+1,\alpha} - h \sum_{r,\alpha} S^z_{r,\alpha};
\\
H_c = -t_\parallel \sum_{r,\alpha} (\psi^\dagger_{r,\alpha}\psi_{r+1,\alpha}+\psi^\dagger_{r+1,\alpha}\psi_{r,\alpha})
-t_\perp\sum_{r} (\psi^\dagger_{r,1}\psi_{r,2}+\psi^\dagger_{r,2}\psi_{r,1})
-\mu\sum_{r,\alpha} \psi^\dagger_{r,\alpha}\psi_{r,\alpha};
\\
H_K= J_K\sum_{r,\alpha,\beta=x,y,z} \psi^\dagger_{r,\alpha}\frac{\sigma^\beta}{2}\psi_{r,\alpha}{\bf S}^\beta_{r,\alpha},
\end{gathered}
\label{eq:hinit}
\end{equation}
where $\psi_{r,\alpha} = (\psi_{r,\alpha,\uparrow},\psi_{r,\alpha,\downarrow})$, $\mu=0$ corresponds to half-filling, and we focus on  $J_\perp \gg J_\parallel$. We first discuss the mapping of the Heisenberg ladder model $H_f$ to an XXZ spin chain and hardcore bosons or Jordan-Wigner fermions for clarity to the reader, as these results are well known \cite{giamarchi.1999}. We then show how to apply this mapping to the Kondo interaction.

Let us for the moment ignore $J_\parallel$ and analyze the resulting states. The ground state (at $h=0$) consists of singlets ($[|\uparrow_{i,1}\downarrow_{i,2}\rangle-|\downarrow_{i,1}\uparrow_{i,2}\rangle]/\sqrt{2}]$) on every rung of the ladder with energy $-3J_\perp/4$. The excited states are the three triplet states on each rung. For $h=0$ the three triplet states are degenerate with the energy $J_\perp/4$. Whereas, for $h\neq0$ the triplet states split and the energies are given by $J_\perp/4\pm h$ and $J_\perp/4$ for $S_z=\pm1$ and $0$, respectively. Under an applied magnetic field, the energy of the lowest triplet state will eventually cross that of the singlet state and a transition occurs.

We can map the two lowest lying states onto a spin-1/2 chain identifying the singlet rung state $(|\uparrow_{i,1}\downarrow_{i,2}\rangle-|\downarrow_{i,1}\uparrow_{i,2}\rangle/\sqrt{2})$ with spin down $|\tilde{\downarrow}\rangle$ and first excited triplet state $|\uparrow_{i,1}\uparrow_{i,2}\rangle$ with spin up $|\tilde{\uparrow}\rangle$. We can now also reintroduce $J_\parallel$ to the model and calculate the matrix elements of the $J_\parallel$ part of Hamiltonian. For this we consider the state of two adjacent rungs:
\begin{gather*}
H_\parallel = J_\parallel ({\bf S}_{a,1} {\bf S}_{b,1}+{\bf S}_{a,2} {\bf S}_{b,2})=
J_\parallel \left(S^z_{a,1}S^z_{b,1}+\frac{1}{2}(S^+_{a,1}S^-_{b,1}+S^-_{a,1}S^+_{b,1})+[1\rightarrow2]\right),
\\
H_\parallel|\tilde{\uparrow}\rangle_a|\tilde{\uparrow}\rangle_b =
\frac{J_\parallel}{2}|\tilde{\uparrow}\rangle_a|\tilde{\uparrow}\rangle_b ,
\\
H_\parallel|\tilde{\uparrow}\rangle_a|\tilde{\downarrow}\rangle_b=
\frac{J_\parallel}{2} |\tilde{\downarrow}\rangle_a|\tilde{\uparrow}\rangle_b,
\\
H_\parallel|\tilde{\downarrow}\rangle_a|\tilde{\downarrow}\rangle_b =
\frac{J_\parallel}{2} |0\rangle_a|0\rangle_b
-J_\parallel |-1\rangle_a|\tilde{\uparrow}\rangle_b
-J_\parallel |\tilde{\uparrow}\rangle_a|-1\rangle_b,
\end{gather*}
where $|-1\rangle \equiv |\downarrow\downarrow\rangle,\;|0\rangle \equiv (|\downarrow\uparrow\rangle +|\uparrow\downarrow\rangle)/\sqrt{2}$ are the higher energy triplet states. If we limit ourselves to the two lowest lying states of each rung we obtain an effective spin-1/2 XXZ Hamiltonian~\cite{giamarchi.1999}:

\begin{equation}
\begin{gathered}
\tilde{H}_f =J_\parallel\sum_i\left(\tilde{S}^x_i\tilde{S}^x_{i+1}+\tilde{S}^y_i\tilde{S}^y_{i+1}+\frac{1}{2}\tilde{S}^z_i\tilde{S}^z_{i+1}\right)
-\tilde{h}\sum_i\tilde{S}^z_i+C,
\\
\tilde{h} = h-J_\perp-J_{\parallel}/2,\;C=(-J_\perp/4-h/2+J_\parallel/8)N.
\end{gathered}
\label{eq:hspin}
\end{equation}

The Hamiltonian Eq. (\ref{eq:hspin}) has a fully polarized (gapped) ground state for $|\tilde{h}|>3J_\parallel/2$ and a gapless phase that can be readily studied by bosonization otherwise. Thus the hapless phase is limited to fields
\begin{equation}
h_c< h <h_c^{up}; \; h_c = J_\perp-J_{\parallel}; \;  h_c^{up} = J_\perp+2J_{\parallel}.
\end{equation}
In this work we will concentrate mostly on the first critical field $h_c$.

Alternatively to the above, the spins can be represented in terms of hard-core bosons using the correspondence
\[
S^z_i=(a_i^\dagger a_i-1/2),\; S^+ = a_i^\dagger, \;S^- = a_i,
\]
leading to the Hamiltonian:
\begin{equation}
\begin{gathered}
H_f^{bos} =
-(J_\parallel/2+\tilde{h})\sum_i a_i^\dagger a_i
+
\frac{J_\parallel}{2}\sum_i(a^\dagger_ia_{i+1}+a^\dagger_{i+1}a_i)
+
\frac{J_\parallel}{2}\sum_i
a_i^\dagger a_i a_{i+1}^\dagger a_{i+1}
+C^b,
\\
C^b=-3J_\perp N/4.
\end{gathered}
\label{eq:hspinbos}
\end{equation}
From this representation it can be seen that the transition is of BEC type, with bosons with momentum $\pi/a$ condensing at the critical field. On the other hand, the spin chain Hamiltonian can be written as well using fermionic (spinon) operators with the help of Jordan-Wigner transformation:
\[
S_i^+\to f_i^\dagger \frac{1}{2} \left[e^{i\pi \sum_{j<i}f_j^\dagger f_j}+e^{-i\pi \sum_{j<i}f_j^\dagger f_j}\right],
S_z = f_i^\dagger f_i - 1/2,
\]
where the string operator is chosen to make the bosonized expressions for spin operators (see below) more convenient.
In the literature, an additional transformation $f_i\to(-1)^if_i$ is usually applied. This results in the following hamiltonian:
\begin{equation}
\begin{gathered}
H_f^{fer} =
-(J_\parallel/2+\tilde{h})\sum_i f_i^\dagger f_i
-\frac{J_\parallel}{2}\sum_i[f^\dagger_{i+1}f_i+f^\dagger_if_{i+1}]
+\frac{J_\parallel}{2}\sum_i f^\dagger_{i+1}f_{i+1}f^\dagger_i f_i
+C^b.
\end{gathered}
\label{eq:hspinfer}
\end{equation}

We move on to the Kondo coupling, that can be rewritten as (to simplify the notation let us consider a single rung of the ladder):
\[
J_K[s_1^z S_1^z + s_2^z S_2^z]
+\frac{J_K}{2} [s_1^+ S_1^-+s_1^- S_1^+ + s_2^+ S_2^- + s_2^- S_2^+],
\]

\begin{gather*}
S_{1(2)}^z= (-)\frac{1}{2}(|0\rangle\langle\tilde{\downarrow}|+|\tilde{\downarrow}\rangle\langle 0|)
+\frac{1}{2}(|\tilde{\uparrow}\rangle\langle\tilde{\uparrow}|-|-1\rangle\langle-1|),
\\
S_{1(2)}^+= -(+) \frac{1}{\sqrt{2}}|\tilde{\uparrow}\rangle\langle\tilde{\downarrow}|
+0|\tilde{\uparrow}\rangle\langle\tilde{\uparrow}|
+\frac{1}{\sqrt{2}}(|0\rangle+(-)|\tilde{\downarrow}\rangle)\langle-1|
+\frac{1}{\sqrt{2}}|\tilde{\uparrow}\rangle\langle0|,
\\
S_{1(2)}^-= +(-) \frac{1}{\sqrt{2}}|-1\rangle\langle\tilde{\downarrow}|
+0|-1\rangle\langle-1|
+\frac{1}{\sqrt{2}}(|0\rangle-(+)|\tilde{\downarrow}\rangle)\langle\tilde{\uparrow}|
+\frac{1}{\sqrt{2}}|\tilde{-1}\rangle\langle0|.
\end{gather*}

Here we consider the limit of small Kondo coupling. In that case, since states $|0\rangle$ and $|-1\rangle$ are always gapped with the energy being of the order $J_\perp$, one can see that the contributions from coupling to this states will be at least of the order $J_K^2/J_\perp$ (using e.g. second-order perturbation theory) and we can neglect them. The remaining matrix elements can be written using the $\tilde{S}^i$ or $a_i$ operators resulting in:
\begin{equation}
\begin{gathered}
H_K \approx \frac{J_K}{2} \sum_i(\tilde{S}^z_i+1/2) (s_{1i}^z+s_{2i}^z)
-\frac{J_K}{2\sqrt{2}} \sum_i\tilde{S}^-_i(s_{1i}^+-s_{2i}^+)+\tilde{S}^+_i(s_{1i}^--s_{2i}^-)
\\
s^\pm_{ai} = c^\dagger_{ai} \sigma^\pm c_{ai},\;s^{x,y,z}_{ai} = c^\dagger_{ai} \frac{\sigma_{x,y,z}}{2} c_{ai},
\end{gathered}
\label{eq:hkproj}
\end{equation}
where introducing $\psi^\pm_i \equiv (c_{1i}\pm c_{2i})/\sqrt{2}$ one gets:
\begin{equation}
\begin{gathered}
H_K \approx \frac{J_K}{4} \sum_i(\tilde{S}^z_i+1/2) (\psi^\dagger_{i,+}\sigma_z\psi_{i,+}+\psi^\dagger_{i,-}\sigma_z\psi_{i,-})-
\\
-\frac{J_K}{2\sqrt{2}} \sum_i\tilde{S}^-_i(\psi^\dagger_{i,+}\sigma^+\psi_{i,-}+\psi^\dagger_{i,-}\sigma^+\psi_{i,+})+
\tilde{S}^+_i(\psi^\dagger_{i,+}\sigma^-\psi_{i,-}+\psi^\dagger_{i,-}\sigma^-\psi_{i,+})=
\\
\frac{J_K}{4} \sum_i a^\dagger_i a_i(\psi^\dagger_{i,+}\sigma_z\psi_{i,+}+\psi^\dagger_{i,-}\sigma_z\psi_{i,-})-
\\
-\frac{J_K}{2\sqrt{2}} \sum_i a_i(\psi^\dagger_{i,+}\sigma^+\psi_{i,-}+\psi^\dagger_{i,-}\sigma^+\psi_{i,+})+
a^\dagger_i(\psi^\dagger_{i,+}\sigma^-\psi_{i,-}+\psi^\dagger_{i,-}\sigma^-\psi_{i,+})
\end{gathered}
\label{eq:hk}
\end{equation}

The Hamiltonian of the conduction electrons can be rewritten as:
\begin{equation}
\begin{gathered}
H_c = \sum_{i,a=\pm}-t_\parallel(\psi^\dagger_{i,a}\psi_{i+1,a}+\psi^\dagger_{i+1,a}\psi_{i,a})
-(\mu+a t_\perp)\psi^\dagger_{i,a}\psi_{i,a}
\\
=
\sum_k \psi^\dagger_{k,+}(\xi_k-t_\perp)\psi_{k,+}+\psi^\dagger_{k,-}(\xi_k+t_\perp)\psi_{k,-},
\end{gathered}
\label{eq:hc}
\end{equation}
where $\xi_k = -2 t_\parallel \cos{k} - \mu$, where we have set the lattice constant to unity for brevity.

\section{One conduction band crossing the Fermi energy}

\subsection{Perturbative calculation for \texorpdfstring{$h<h_c$}{}}
In the regime $h<h_c(J_K)$ the boson density is assumed to be zero. As the Hamiltonian is quadratic in fermionic fields we proceed by integrating them out and then expanding in powers of the bosonic fields. Here we ignore the hard-core constraint and work with usual Bose fields; formally one may replace the constraint by an on-site repulsion that is much larger then the bandwidth. Since we are interested in qualitative effects due to the Kondo coupling, we may examine the terms in the action generated by the Kondo coupling in the free boson limit and then study the effect of these new terms if a hard-core constraint is implemented. The partition function reads
\begin{gather*}
Z = \int D\overline{\psi}_+ D\psi_+ D\overline{\psi}_- D\psi_- D a^\dagger D a
\exp\left\{-S_f^{bos}[a^\dagger,a]-S_c[\overline{\psi}_+ ,\psi_+,\overline{\psi}_- ,\psi_-]-S_K\right\}=
\\
Z_c \int D a^\dagger D a
\exp\left\{-\left[S_f^{bos}[a^\dagger,a]-\log[\langle e^{-S_K} \rangle_c]\right]\right\}=
\\
Z_c \int D a^\dagger D a
\exp\left\{-\left[S_f^{bos}[a^\dagger,a]- (\langle e^{-S_K} \rangle_c^{conn}-1)\right]\right\}.
\end{gather*}
where $\langle A \rangle_c$ denotes the expectation value of an arbitrary operator $A$ and is given by
\begin{equation}
\langle A \rangle_c = \frac{\int D\overline{\psi}_+ D\psi_+ D\overline{\psi}_- D\psi_- A e^{-S_c}}
{\int D\overline{\psi}_+ D\psi_+ D\overline{\psi}_- D\psi_-  e^{-S_c}}.
\end{equation}
Moreover, as the bosons are dilute it makes sense to decompose the induced term in powers of $a/a^\dagger$ fields. There are two types of couplings that correspond to the first and second term in Eq. (\ref{eq:hk}) represented as vertices in Fig. \ref{fig2}.

\begin{figure}[h!]
	\centering
	\includegraphics[width=0.8\linewidth]{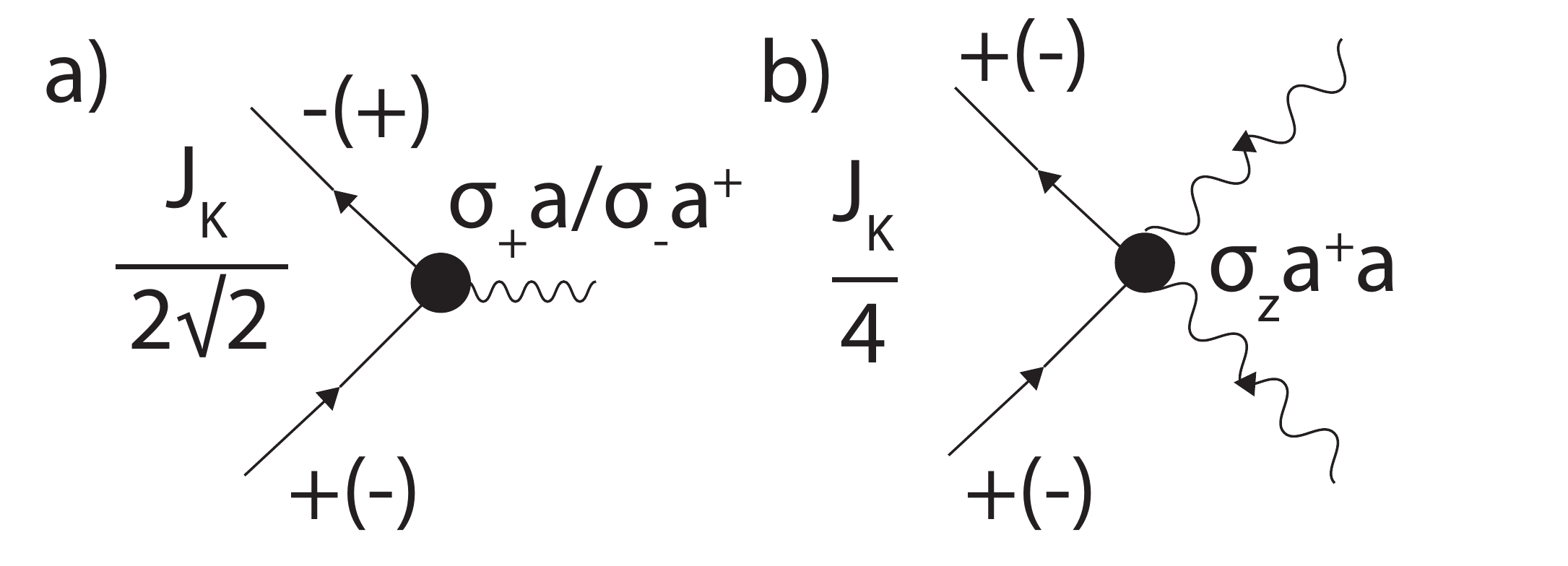}
	\caption{Interactions between the hard-core bosons and conduction electrons stemming from a) second b) first terms in Eq.  (\ref{eq:hk})}
	\label{fig2}
\end{figure}

\subsubsection{Lowest order}
At the lowest order in bosonic fields $a/a^\dagger$ there is only one diagram that contributes to the effective action (see Fig. \ref{fig3} a)). The expression is (note that the minus signs due to the fermionic bubble in the diagram and reexponentiating cancel):
\begin{gather*}
\delta S_{bos}
=\int \frac{dq}{2\pi} a^\dagger_q a_q \int\frac{d\omega}{2\pi}\frac{dk}{2\pi}
\frac{J_K^2}{8}
\left(
\frac{1}{(i\omega-\xi_k +t_\perp)(i\omega - \xi_{k+q}-t_\perp)}
+\frac{1}{(i\omega-\xi_k -t_\perp)(i\omega - \xi_{k+q}+t_\perp)}
\right)=
\\
=
\int \frac{dq}{2\pi} a^\dagger_q a_q I_{bos}(q).
\end{gather*}
\begin{figure}[h!]
	\centering
	\includegraphics[width=0.8\linewidth]{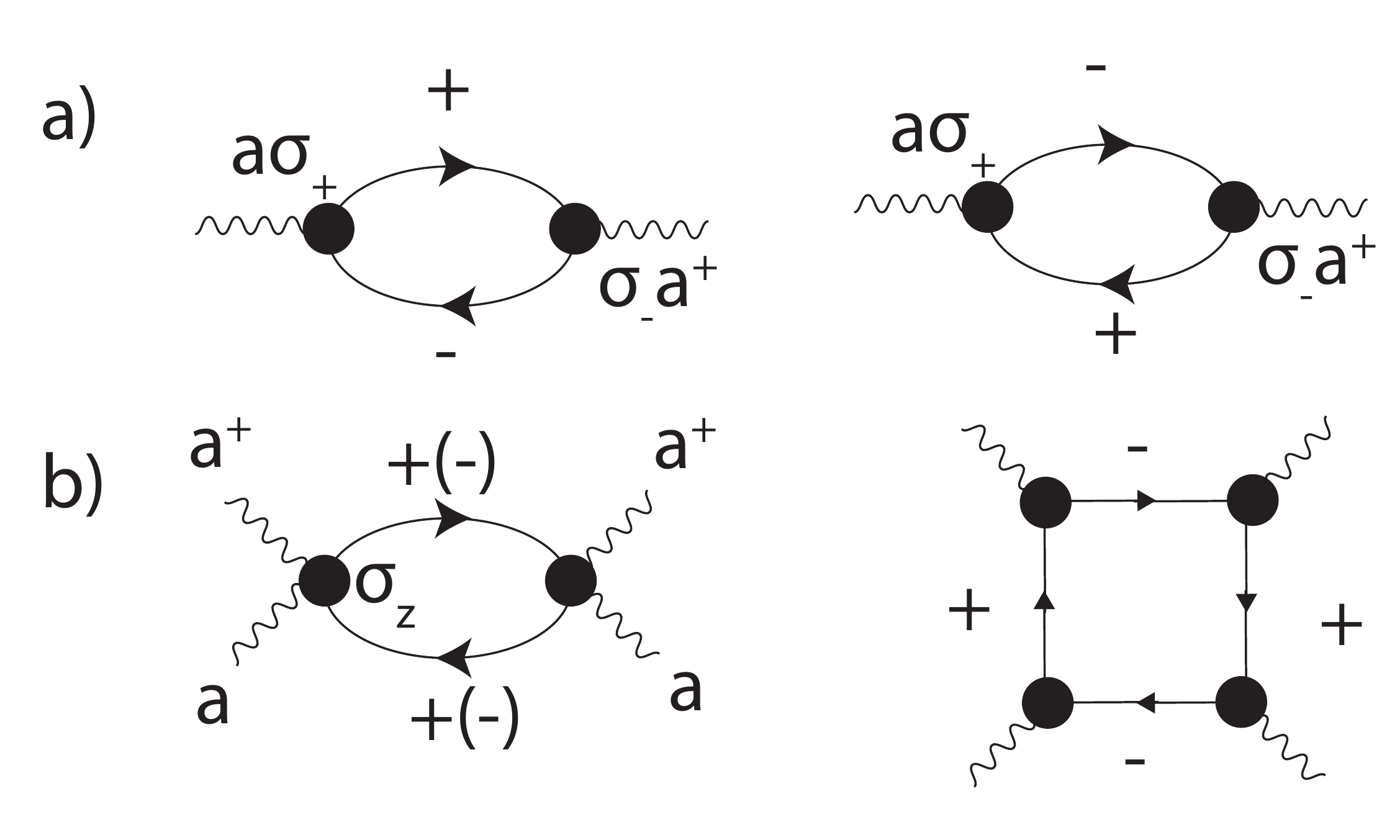}
	\caption{Diagrams contributing to the effective action a) lowest order in $a/a^\dagger$ b) fourth order in $a/a^\dagger$.}
	\label{fig3}
\end{figure}
First of all it is easy to see that it's negative
\[
I_{bos}(q)\sim\int \frac{d\omega}{2\pi} \frac{1}{(i\omega-a)(i\omega - b)}=
-\frac{\theta(a)\theta(-b)}{a-b}-\frac{\theta(-a)\theta(b)}{b-a}<0,
\]
such that the critical field for the BEC is {\it lower} in the presence of a small $J_K$. Note that if the g-factor of itinerant electrons had not been neglected there would have been an additional contribution $\sim - J_K g_c h v_F a^\dagger a/2\pi$ due to the $ S^z_i\cdot  s^z_i$ part of the Kondo coupling (referred to as the ZZ part hereafter) further decreasing $h_{c}(J_K)$. Now let us discuss the value of the present contribution. For the case Fig. 2 a),b) of the main text only one of the bands can be present at the Fermi level and consequently the integral is cut off at low energies and one expects controlled behavior. First we rewrite the integral using dimensionless parameters:
\begin{equation}
\begin{gathered}
I_{bos}(q)=-\frac{J_K^2}{8 t_\perp}
\int_{0}^{k_F}\frac{dx}{2\pi}
\left(
\frac{1}{1+2\frac{t_\parallel}{t_\perp}\sin\frac{q}{2}\sin\left(x+q/2\right)}
+\frac{1}{1-2\frac{t_\parallel}{t_\perp}\sin\frac{q}{2}\sin\left(x-q/2\right)}
\right), \; \nu<1/2
\\
\,\,\,\,\,\,\,\,\,\,\,\,\,\,\,\,\,
=-\frac{J_K^2}{8 t_\perp}
\int_{k_F}^\pi\frac{dx}{2\pi}
\left(
\frac{1}{1-2\frac{t_\parallel}{t_\perp}\sin\frac{q}{2}\sin\left(x+q/2\right)}
+\frac{1}{1+2\frac{t_\parallel}{t_\perp}\sin\frac{q}{2}\sin\left(x-q/2\right)}
\right), \; \nu>1/2
\end{gathered}
\label{eq:IndDisp}
\end{equation}
The results are presented in Fig. \ref{fig:IndDisp}. One can see that in the large part of the phase space the maximum is reached at $q=0$ (the value is then just $-\frac{J_K^2}{4 t_\perp} \nu$ for $\nu<1/2$ and $-\frac{J_K^2}{4 t_\perp}(1-\nu)$ for $\nu>1/2$). However, the difference between the value at maximum and at $q=\pi$ (where the original spinon dispersion has minimum) is of order 1. As we assume $J_K\ll J_\parallel, t_\perp$ it follows that the renormalized spinon dispersion still has a minimum at $q=\pi$.
\begin{figure}[h!]
	\centering
	\includegraphics[width=0.8\linewidth]{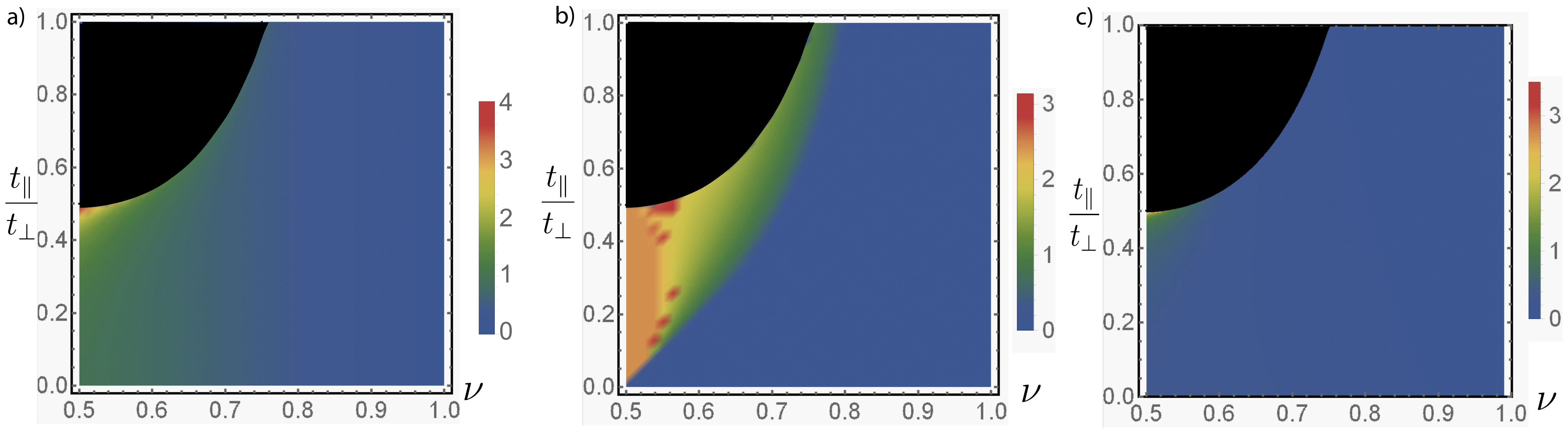}
	\caption{ a) Maximal value of the dimensionless integral in Eq. (\ref{eq:IndDisp}) as a function of $q$ b) value of $q$ (from $0$ to $\pi$) maximizing Eq. (\ref{eq:IndDisp}) c) value of the integral in Eq. (\ref{eq:IndDisp}) for $q=\pi$. The excluded black region is where the second band crosses the Fermi level.}
	\label{fig:IndDisp}
\end{figure}

On the other hand, for $t_\perp\leq 2 t_\parallel, \mu<-t_\perp+2t_\parallel$ (Fig. 2 c)) the integral has a singularity at $q=k_F^+ +k_F^-$ resembling that of the Peierls transition. Analysis of this case is presented in the next section.
\subsubsection{Higher orders}
If we now consider the next order diagrams, a peculiar result is that the Peierls singularity is seen in the first diagram of Fig. \ref{fig3} b) at momentum transfer $q=2k_F$. However, since there are no low-energy bosons to take advantage of this interaction the system is stable.

However, the same diagram results in a non-zero self energy for the bosonic propagator depicted in Fig \ref{pol2}. The expression corresponding to the diagram is:
\begin{equation}
2\frac{J_K^2}{16}\int \frac{d\omega dq}{(2\pi)^2}\frac{d\varepsilon dp}{(2\pi)^2}
\frac{1}{i\omega - \varepsilon_{q+q_0}} \frac{1}{i(\varepsilon-\omega) - \xi^\pm_{k-q}}
\frac{1}{i\varepsilon - \xi^\pm_{k}},
\end{equation}
where $2$ in front is due to summation over spin indices, $q_0$ is the incoming momentum and we set the incoming frequency to be zero, as $h_c$ is set by $D^{-1}(i\omega=0,q_0) = 0$, $D$ being the bosonic propagator. For the one -band case when the "+" band is partially occupied while the "-" band is empty, the latter's contribution vanishes (e.g. the integral over $\varepsilon$ is zero as the poles are in the same half-plane), while the former is:
\begin{equation}
\begin{gathered}
i\frac{J_K^2}{8}\int \frac{d\omega dq}{(2\pi)^2}\frac{d\varepsilon dp}{(2\pi)^2}
\frac{1}{\omega + i \varepsilon_{q+q_0}} \frac{1}{\varepsilon-\omega +i \xi^+_{k}}
\frac{1}{\varepsilon +i \xi^+_{k+q}}=
\\
=
\frac{J_K^2}{8}\int \frac{dq}{2\pi}\frac{d\varepsilon dp}{(2\pi)^2}
\frac{\theta(\xi^+_{k})}{\varepsilon+i \varepsilon_{q+q_0}+i \xi^+_{k}}
\frac{1}{\varepsilon +i \xi^+_{k+q}}=
\\
\frac{J_K^2}{8}\int \frac{dq}{2\pi}\frac{dp}{2\pi}
\frac{\theta(\xi^+_{k})\theta(-\xi^+_{k+q})}{\varepsilon_{q+q_0}+ \xi^+_{k}-\xi^+_{k+q}}.
\end{gathered}
\end{equation}
One can see that even if $\varepsilon_{q+q_0} = 0$ the denominator is nonzero for all the integration region; thus we can take $q_0=\pi,\varepsilon_{q+q_0} = J_\parallel (-\cos q+1);\;\xi^+_{k}-\xi^+_{k+q} = -2t_\parallel[\cos (k) - \cos (k+q)]$. One can rewrite the integral as:
\begin{equation}
\delta h_c^{(2)} = -\frac{J_K^2}{8}\int_{|k|>k_F} \frac{dk}{2\pi}\int_{-k_F}^{k_F}\frac{dk'}{2\pi} \frac{1}{J_\parallel( -\cos(k'-k)+1)-2t_\parallel[\cos (k) - \cos (k')]},
\label{eq:hc2cor}
\end{equation}
For 1/8 filling $k_F=\pi/4$, for $J_\parallel = 0.3, t_\parallel = 1$, one gets $\delta h_c^{(2)}\approx-0.0129 J_K^2$, while the contribution of the first diagram \eqref{eq:IndDisp} is $\approx -0.009 J_K^2$. The total result is then $\delta h_c \approx -0.022 J_K^2$.

\begin{figure}[h!]
	\centering
	\includegraphics[width=0.5\linewidth]{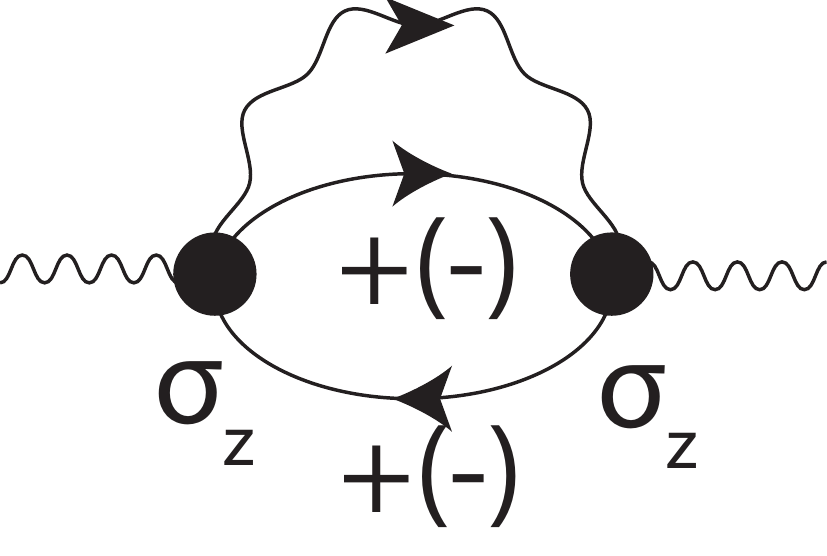}
	\caption{Bosonic self-energy from induced interaction in Fig. \ref{fig3} b).  }
	\label{pol2}
\end{figure}

We attribute the quantitative discrepancy between these results and the numerical ones to the effect of the hard-core constraint. In particular, while the contribution \eqref{eq:hc2cor}, if derived with conventional perturbation theory, would contain no contributions from virtual two-boson states (because it stems from the ZZ part of the Kondo coupling, Eq. [2] of the main text, that manifestly conserves boson number), the expression \eqref{eq:IndDisp} does contain one (because it stems from the XY part of the Kondo coupling, Eq. [2] of the main text, that allows for transitions between 1-boson and 2-boson states). The effect of the constraint is most easily observed in the local limit $t_\parallel= J_\parallel=0$. In that case we may consider a single site and study the perturbative corrections to the energies of the zero-boson and one-boson states due to the second line of Eq. [2] of the main text, their difference at $h-h_c^0$ being the correction to $h_c$ from this term. We also assume the lower "+" band to be completely filled (occupied with two electrons for a single site). For ordinary bosons, virtual transitions from (0) to (1) leads to a shift of $-\frac{J_K^2}{16 t_\perp}$ of the energy of the (0) states, while from the (1) one can either go to (0), leading to an $-\frac{J_K^2}{16 t_\perp}$ shift or (2) leading to an additional  $-\frac{2J_K^2}{16 t_\perp}$, where the factor of 2 is due to the bosonic creation operator matrix element. Overall, this leads to $\delta h_c^{free} = -\frac{2J_K^2}{16 t_\perp}$. In the hard-core limit, on the other hand, the transition (1)$\to$(2) is forbidden and thus $\delta h_c^{h-c} = 0$. Thus, the contribution in \eqref{eq:IndDisp} can be strongly suppressed by the hard-core constraint, partially explaining the quantitaive discrepancy between Eq. [3] of the main text and the DMRG result. We also note that contributions from higher triplet states with $S_z=0,-1$ may additionally affect $h_c$; however, since these states are gapped, the functional form $h_c(J_K=0)-h_c(J_K)\sim J_K^2$ may not be affected by these contributions.

\subsubsection{Effect of the interaction on the fermions}

We can also look at the effective low-energy properties of the fermions due to integrating the bosons out. Using perturbation theory one gets the following interaction term:
\[
S_{ind}^{h<h_c} = -\frac{J_K^2}{32} \int dx d \tau \int dx' d\tau'\langle T_\tau (\tilde{S}_z(x,\tau)+1/2) (\tilde{S}_z(x',\tau')+1/2)\rangle  [\psi_\pm^\dagger(x,\tau)\sigma_z \psi_\pm(x,\tau)][\psi_\pm^\dagger(x',\tau')\sigma_z \psi_\pm(x',\tau')].
\]
Using the fermionic expressions for spins one however has:
\[
S_{ind}^{h<h_c}\sim\langle T_\tau (\tilde{S}_z(x)+1/2) (\tilde{S}_z(x')+1/2)\rangle = \langle T_\tau f^\dagger(x,\tau)f(x,\tau) f^\dagger(x',\tau')f(x',\tau')\rangle
= -G_f(x-x',\tau-\tau')G_f(x'-x,\tau'-\tau)=0,
\]
where $G_f(x-x',\tau-\tau') = -\langle T_\tau f(x,\tau) f^\dagger(x',\tau')\rangle$ and Wick's theorem was used; averaging in the Green's functions is carried over the vacuum state. Hence, the fermions are not affected in this case for fields below the critical one.

Let us now consider the XX and YY couplings. As they include the gapped fermions it is a good idea to get rid of them first. We have the action:

\begin{gather*}
S_0^++S_{K}^{XY} = \sum_{i,i'}\psi^\dagger_{i,+} (-G_0^{-1})_{ii'} \psi_{i',+}
+
\frac{J_K}{2\sqrt{2}} \sum_i \tilde{S}^-_i(\psi^\dagger_{i,+}\sigma^+\psi_{i,-}+\psi^\dagger_{i,-}\sigma^+\psi_{i,+})+
\tilde{S}^+_i(\psi^\dagger_{i,+}\sigma^-\psi_{i,-}+\psi^\dagger_{i,-}\sigma^-\psi_{i,+})=
\\
\sum_{i,i'}
(\psi^\dagger_{+,\uparrow} - \frac{J_K}{2\sqrt{2}}\tilde{S}^+ \psi^\dagger_{-,\downarrow} G_0)_i
(-G_0^{-1})_{ii'}
(\psi_{+,\uparrow} - G_0 \frac{J_K}{2\sqrt{2}}\tilde{S}^-\psi_{-,\downarrow})_{i'}
\\
+
(\psi^\dagger_{+,\downarrow} - \frac{J_K}{2\sqrt{2}}\tilde{S}^-  \psi^\dagger_{-,\uparrow} G_0)_i
(-G_0^{-1})_{ii'}
(\psi_{+,\downarrow} - G_0 \frac{J_K}{2\sqrt{2}}\tilde{S}^+\psi_{-,\uparrow})_{i'}
\\
+
\frac{J_K^2}{8}\tilde{S}^+_i\tilde{S}^-_{i'} \psi^\dagger_{-,i,\downarrow} G_0 \psi_{-,i',\downarrow}
+
\frac{J_K^2}{8}\tilde{S}^-_i\tilde{S}^+_{i'} \psi^\dagger_{-,i,\uparrow} G_0 \psi_{-,i',\uparrow},
\end{gather*}
where $(G_0)_{ii}$ is the Fourier transform of
\begin{equation}
G_0(i \varepsilon,k) = \frac{1}{i \varepsilon - \xi_+(k)}.
\label{eq:G0}
\end{equation}
If we omit the frequency dependence of $G_0(i \varepsilon,k)$ and take the extreme case $t_\perp\gg t_\parallel$ the last two terms take the form
\[
\sum_i
\frac{J_K^2}{16 t_\perp}\psi^\dagger_{i,-} \psi_{i,-}
-
\frac{J_K^2}{8 t_\perp}\tilde{S}^z_i\psi^\dagger_{i,-} \sigma_z \psi_{i,-},
\]
which is the ZZ coupling already considered (the first term is absorbed in the conduction electrons' chemical potential). Furthermore, this coupling is smaller then the original ZZ one by the parameter $J_K/ t_\perp$ and thus can be neglected in the low-energy theory.

However, there still may be effects that arise from the high frequencies and where the frequency dependence of $G_0$ can not be omitted. We turn now to consideration of such effects by considering lowest order diagrams.

\subsubsection{Self-energy and magnetization} 
\label{sec:selfen}

\begin{figure}[h!]
	\centering
	\includegraphics[width=0.8\linewidth]{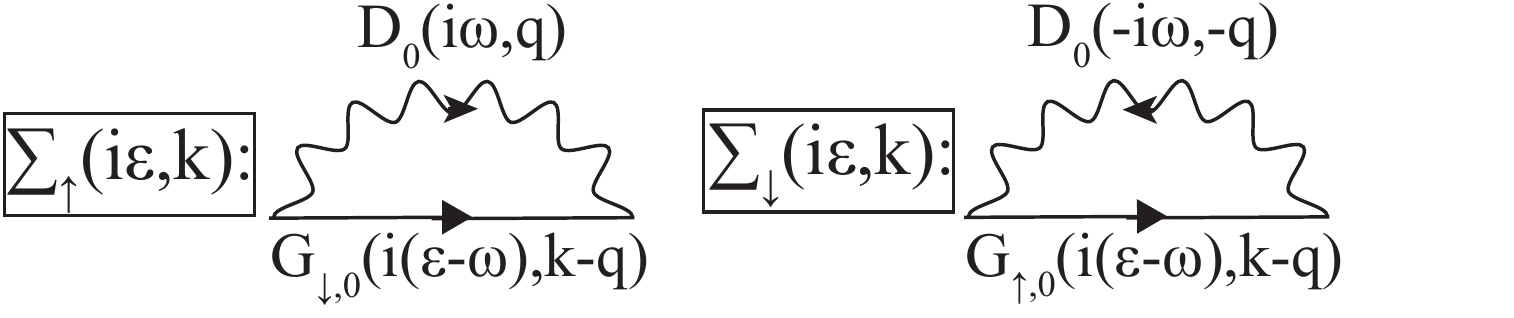}
	\caption{Self-energy diagrams for fermions (band index suppressed); interaction vertices defined in Fig. \ref{fig2}}
	\label{fig:selfen}
\end{figure}

Let us first study the lowest-order diagrams contributing to the fermionic self-energy (see Fig. \ref{fig:selfen}). $G_0$ is defined by Eq. \eqref{eq:G0} and the bosonic propagator $D_0$ is:
\begin{equation}
D_0(i\omega,q) = \frac{1}{i \omega- \varepsilon_q},
\label{eq:D0}
\end{equation}
where $\varepsilon_q = J_\perp+ J_\parallel \cos q - h$ is the bosonic dispersion relation from Eq. \eqref{eq:hspinbos}. The explicit expressions for the self-energies are:
\begin{gather*}
\Sigma_{+(-),\uparrow} (i\varepsilon, k) = -\frac{J_K^2}{8} \int \frac{d\omega}{2\pi}  \frac{ d q}{2\pi}
\frac{1}{i \omega - \varepsilon_q} \frac{1}{i(\varepsilon - \omega) - \xi^{-(+)}_{k+q}}
=
-\frac{J_K^2}{8} \int
\frac{ d q}{2\pi}  \frac{i \theta[\xi^{-(+)}_{k+q}] }{\varepsilon+i\xi^{-(+)}_{k+q}+i\varepsilon_q};
\\
\Sigma_{+(-),\downarrow} (i\varepsilon, k) = -\frac{J_K^2}{8} \int \frac{d\omega}{2\pi}  \frac{ d q}{2\pi}
\frac{1}{-i \omega - \varepsilon_q} \frac{1}{i(\varepsilon - \omega) - \xi^{-(+)}_{k+q}}
=
-\frac{J_K^2}{8} \int
\frac{ d q}{2\pi}  \frac{i \theta[-\xi^{-(+)}_{k+q}] }{\varepsilon+i\xi^{-(+)}_{k+q}-i\varepsilon_q},
\end{gather*}
where $\xi^{-(+)}_{k+q} = \pm t_\perp+\xi_k = \pm t_\perp-2 t_\parallel \cos{k} - \mu$. Or, in real frequencies after a Wick rotation:
\begin{gather*}
\Sigma_{+(-),\uparrow} (\varepsilon, k) =
-\frac{J_K^2}{8} \int \frac{ d q}{2\pi}
\frac{\theta[\xi^{-(+)}_{k+q}] }{\xi^{-(+)}_{k+q}+\varepsilon_q-\varepsilon-i \delta};
\\
\Sigma_{+(-),\downarrow} (\varepsilon, k) =
\frac{J_K^2}{8} \int
\frac{ d q}{2\pi}  \frac{\theta[-\xi^{-(+)}_{k+q}] }{\varepsilon+\varepsilon_q-\xi^{-(+)}_{k+q} -i\delta};
\end{gather*}

Most importantly, the self energies for spin up and down electrons are not equal to each other leading to a Zeeman effect. Namely, close to the Fermi level we can take $\varepsilon\approx 0$ to get:

\begin{equation}
\Sigma_{+(-),\uparrow} (0, k_F)  - \Sigma_{+(-),\downarrow} (0, k_F)
=
-\frac{J_K^2}{8} \int \frac{ d q}{2\pi}
\frac{1}{\varepsilon_q+|\xi^{-(+)}_{k_F+q}|},
\label{eq:selfenZ}
\end{equation}
where the $i\delta$ can be dropped since the denominators are never zero in the gapped phase. Note that this difference can be backtracked to the frequency dependence of the bosonic propagator. Additionally, unless $k_F=\pi/2$ (i.e. one of the bands is half filled), this quantity is not singular at the BEC transition. This result suggests that the energy of spin up electrons is smaller then that of spin down ones, which results (at a fixed chemical potential) in a depopulation of the spin down subband with respect to spin down one. This also suggests that a) a splitting of the Fermi points for up/down electrons even if $g_c = 0$; b) finite magnetization of the conduction electrons.

We can also check the latter statement directly. Let us now calculate the magnetization of conduction electrons and localized spins perturbatively (i.e. the correction to Green's functions is taken in the lowest order in the interaction):
\begin{gather*}
2*M^z_{c} = \langle \psi_{+,\uparrow}^\dagger\psi_{+,\uparrow} -  \psi_{+,\downarrow}^\dagger\psi_{+,\downarrow}\rangle
+\langle \psi_{-,\uparrow}^\dagger\psi_{-,\uparrow} -  \psi_{-,\downarrow}^\dagger\psi_{-,\downarrow} \rangle =
\\
\int \frac{d \varepsilon d k}{(2\pi)^2}[ G_{\uparrow}^+(i\varepsilon, k) - G^+_{\downarrow}(i\varepsilon, k)+G_{\uparrow}^-(i\varepsilon, k) - G^-_{\downarrow}(i\varepsilon, k)]e^{i\varepsilon\delta} \approx
\\
\approx
-\frac{J_K^2}{8}\int \frac{d \varepsilon d k d\omega d q}{(2\pi)^4}
\{
[G_{0,\uparrow}^+(i\varepsilon, k)]^2 G^+_{0,\downarrow}(i(\varepsilon-\omega), k-q)D_0(i \omega,q)
\\
-[G_{0,\downarrow}^+(i\varepsilon, k)]^2 G^+_{0,\uparrow}(i(\varepsilon-\omega), k-q)D_0(-i \omega,-q)
+ ...[+\leftrightarrow-]
\}.
\end{gather*}
Evaluating the integral over $\omega$ in the first term (using $\varepsilon_q>0$, i.e. bosons are gapped) we get:
\begin{gather*}
I_1=-\int \frac{d \varepsilon  d k d\omega d q}{(2\pi)^4}
\frac{1}{(i\varepsilon - \xi_+^k)^2}\frac{1}{i(\varepsilon-\omega) - \xi_-^{k-q}}
\frac{1}{i\omega - \varepsilon_q}
=
\int \frac{d \varepsilon d k d q}{(2\pi)^3}
\frac{1}{(i\varepsilon - \xi_+^k)^2}\frac{\theta[ \xi_-^{k-q}]}{i\varepsilon-\varepsilon_q - \xi_-^{k+q}}
\\
=
-\int \frac{ d k d q}{(2\pi)^2}
\frac{\theta[\xi_-^{k-q}]\theta[-\xi_+^k]}{(\varepsilon_q + \xi_-^{k+q}- \xi_+^k)^2}
\end{gather*}
Analogously performing the integrals for the second term we get:
\begin{gather*}
I_2=-\int \frac{ d \varepsilon d k d\omega d q}{(2\pi)^4}
\frac{1}{(i\varepsilon - \xi_+^k)^2}\frac{1}{i(\varepsilon-\omega) - \xi_-^{k-q}}
\frac{-1}{i\omega + \varepsilon_q}
=
\int \frac{d \varepsilon  d k d q}{(2\pi)^3}
\frac{1}{(i\varepsilon - \xi_+^k)^2}\frac{\theta[-\xi_-^{k-q}]}{i\varepsilon+\varepsilon_q - \xi_-^{k+q}}
\\
=
\int \frac{ d k d q}{(2\pi)^2}
\frac{\theta[-\xi_-^{k-q}]\theta[\xi_+^k]}{(\varepsilon_q - \xi_-^{k+q}+ \xi_+^k)^2}.
\end{gather*}
Altogether, one gets:

\begin{equation}
\begin{gathered}
M^z_{c} =
-
\frac{1}{2}
\frac{J_K^2}{8}
\int \frac{ d k d q}{(2\pi)^2}
\frac{1}{(\varepsilon_q + |\xi_-^{k-q}|+ |\xi_+^k|)^2}
+
\frac{1}{(\varepsilon_q + |\xi_+^{k-q}|+ |\xi_-^k|)^2} =
\\
=
-
\frac{J_K^2}{8}
\int \frac{ d k d q}{(2\pi)^2}
\frac{1}{(\varepsilon_q + |\xi_-^{k-q}|+ |\xi_+^k|)^2},
\end{gathered}
\label{eq:mcondpert}
\end{equation}
the last line obtained with $k\to k+q$, $q\to -q$ using $\varepsilon_{-q} = \varepsilon_q$. Interestingly, the total magnetization actually turns out to be directed opposite to magnetic field, despite Zeeman splitting lowering the energy of the spins along the field. This can be backtracked to the presence of a pole in the fermionic self energy, which contributes to the magnetization as is shown above.

For the localized spins total magnetization is equal to the number of bosons (note the additional minus sign due to the fermionic bubble in the diagram):
\begin{gather*}
M^z_f = n_B = -\int \frac{d\omega d q}{(2 \pi)^2} D(i\omega,q) e^{i\omega\delta} \approx
\\
\approx
-\frac{J_K^2}{8}\int \frac{ d \varepsilon d k d\omega d q}{(2\pi)^4}
[D_0(i\omega,q)]^2
\{
G_{0,\downarrow}^+(i(\varepsilon-\omega), k-q)G_{0,\uparrow}^-(i\varepsilon, k)
+
G_{0,\downarrow}^-(i(\varepsilon-\omega), k-q)G_{0,\uparrow}^+(i\varepsilon, k)
\}
\\
=
-\frac{J_K^2}{8}\int \frac{ d \varepsilon d k d\omega d q}{(2\pi)^4}
\frac{1}{(i\omega - \varepsilon_q)^2}
\left\{
\frac{1}{i\varepsilon - \xi_+^k}\frac{1}{i(\varepsilon-\omega) - \xi_-^{k-q}}
+
\frac{1}{i\varepsilon - \xi_-^k}\frac{1}{i(\varepsilon-\omega) - \xi_+^{k-q}}
\right\}
=
\\
=
\frac{J_K^2}{8}\int \frac{d \varepsilon  d k d q}{(2\pi)^3}
\frac{\theta[\xi_-^{k-q}]}{(i\varepsilon- \xi_-^{k-q} - \varepsilon_q)^2}
\frac{1}{i\varepsilon - \xi_+^k}
+
\frac{\theta[\xi_+^{k-q}]}{(i\varepsilon- \xi_+^{k-q} - \varepsilon_q)^2}
\frac{1}{i\varepsilon - \xi_-^k}=
\\
=
\frac{J_K^2}{8}\int \frac{ d k d q}{(2\pi)^2}
\frac{\theta[\xi_-^{k-q}]\theta[- \xi_+^k]}{(-\xi_+^k+ \xi_-^{k-q} + \varepsilon_q)^2}
+
\frac{\theta[\xi_+^{k-q}]\theta[- \xi_-^k]}{(-\xi_-^k+ \xi_+^{k-q} + \varepsilon_q)^2},
\end{gather*}
resulting in
\begin{equation}
\begin{gathered}
M^z_{f} =
\frac{J_K^2}{8}
\int \frac{ d k d q}{(2\pi)^2}
\frac{1}{(\varepsilon_q + |\xi_-^{k-q}|+ |\xi_+^k|)^2}.
\end{gathered}
\label{eq:mlocpert}
\end{equation}
It is evident from Eqs. \eqref{eq:mcondpert}, \eqref{eq:mlocpert} that the total magnetization $M_{cond}+M_{loc}$ vanishes; i.e. despite the presence of Zeeman splitting the system is not magnetized. Note that the effect of the ZZ term in the Kondo Hamiltonian is of higher order in $J_K$, as the magnetization for both the conduction electrons and the localized spins is perturbatively small. Once again, at the BEC transition only $q=\pi$ becomes gapless, so unless the fermions are at half-filling, the contribution to magnetization described above will evolve smoothly through $h_{c}$.

Finally, let us remark that the effects and expressions above are valid also for the case where both bands cross the Fermi level. Gap in the bosonic spectrum still guarantees that the corrections obtained are indeed small and no divergencies occur.


\subsection{Bosonization analysis for \texorpdfstring{$h>h_c$}{}}
In this case the hard-core bosons/spinons stemming from the spin ladder are not gapped and we can not ignore the interactions between them or simply integrate them out. We can, however, linearize near the Fermi points of the Jordan-Wigner (spinon) representation of the spin and then use bosonization to include the interactions non-perturbatively. One can also use directly the spin-boson mapping (note that these bosons describe the low-energy degrees of freedom of spinons and are different from the hard-core bosons introduced previously):
\begin{equation}
\tilde{S}_z(x) = -\frac{1}{\pi}\partial_x\varphi+\frac{1}{\pi\tilde\alpha}\cos(2\varphi(x)-2k_F^fx),
\;
\tilde{S}^+(x) = (-1)^x \frac{e^{-i\theta(x)}}{\sqrt{2\pi\tilde\alpha}}[1+\cos(2\varphi(x)-2k_F^fx)],
\label{eq:spinboson}
\end{equation}
where $k_F^f$ is the filling of the Jordan-Wigner fermion band that depends on $\tilde{h} = h-J_\perp-J_\parallel/2$ and $\tilde\alpha$ is the lattice cutoff of the order of the lattice constant. For $\tilde{h}=0$ (halfway through the gapless phase) one should have $\langle \tilde{S}_z \rangle = \langle f^\dagger_if_i -1/2\rangle = 0$ and thus the band is at half filling $k_F^f=\pi$. The fermionic hopping term transforms into the bare term
\[
H_0^f = \frac{1}{2\pi} \int dx v_F [(\partial_x\theta(x))^2+(\partial_x\varphi(x))^2],
\]
where $v_F = J_\parallel \sin (k_F^f a)$. The interaction term, which is due to ZZ coupling is then:
\begin{gather*}
H^f_{int}=\frac{J_\parallel}{2}\sum_x
\left[
-\frac{1}{\pi}\partial_x\varphi(x+a)+\frac{1}{\pi\tilde\alpha}\cos(2\varphi(x+a)-2k_F^f (x+a) )
\right]
\left[
-\frac{1}{\pi}\partial_x\varphi(x)+\frac{1}{\pi\tilde\alpha}\cos(2\varphi(x)-2k_F^f x)
\right]\approx
\\
\approx
\frac{J_\parallel}{2} \int dx\frac{1}{\pi^2}(\partial_x\varphi(x))^2
+
\frac{1}{(\pi\tilde\alpha)^2}\cos(2\varphi(x+a)-2k_F^f (x+a))\cos(2\varphi(x)-2k_F^f x)\approx
\\
\frac{J_\parallel}{2}(1-\cos(2k_F^f a)) \int dx\frac{1}{\pi^2}(\partial_x\varphi(x))^2
+
\frac{1}{2(\pi\tilde\alpha)^2}\cos(4\varphi(x)-4 k_F^f x-2k_F^fa),
\end{gather*}
The last term loses its relevance in a magnetic field away from half filling. Moreover, one can estimate its scaling dimension for $\tilde{h} = 0$: as $\langle e^{i a[\varphi(x)]}e^{-i a[\varphi(0)]} \rangle\sim r^{-a^2 K/2}$ for large $r$, $[\cos(4\varphi(x))] = L^{-4 K}$ which means that it is irrelevant for $K>1/2$. Indeed the exact solution has $K=3/4$ \cite{giamarchi.1999} and only grows with increasing field. Essentially, spins throughout the 'superfluid' phase are described by a LL Hamiltonian:
\begin{equation}
H^f_{SF} = \frac{1}{2\pi} \int dx  u K [(\partial_x\theta(x))^2+\frac{u}{K}(\partial_x \varphi)^2],
\end{equation}
where $\varphi\to\varphi-\pi (M^z_{f} -1/2) x$ in a finite $\tilde{h}$. Note that here the magnetization $M^z_{f}$ is not induced by the Kondo coupling as was discussed above for $h<h_c$. Consequently, in the leading order we will neglect the perturbative corrections to $M^z_{f}$ here. $(M^z_{f} -1/2)$ is given by $K h/u$, note however that $K$ and $u$ are themselves function of $h$.

We now bosonize the itinerant fermions using the expressions for the fermionic operators:
\begin{gather*}
\psi_{r,\sigma}(x)= \frac{1}{\sqrt{2\pi\tilde\alpha}} U_{r,\sigma} e^{irk_Fx}e^{-\frac{i}{\sqrt{2}}[r\varphi_\rho(x)-\theta_\rho(x)+\sigma(r\varphi_\sigma(x)-\theta_\sigma(x))]},
\\
s_z(x) = \rho_\uparrow(x)-\rho_\downarrow(x) =
-\frac{\sqrt{2}\partial_x \varphi_\sigma}{\pi}+
\frac{1}{\pi\tilde\alpha}\left[\cos(2 k_F x - \sqrt{2} (\varphi_\rho+\varphi_\sigma))
-\cos(2 k_F x - \sqrt{2} (\varphi_\rho-\varphi_\sigma))
\right].
\end{gather*}

Now let us consider the Kondo term. We start with the ZZ part:
\begin{gather*}
H_K^{ZZ} = \frac{J_K}{4} \int dx
\left(
-\frac{1}{\pi}\partial_x\varphi(x)+\frac{1}{\pi\tilde\alpha}\cos(2\varphi(x)-2k_F^f x)+M^z_f
\right)
\\
\cdot
\left(
-\frac{\sqrt{2}\partial_x \varphi_\sigma}{\pi}+
\frac{1}{\pi\tilde\alpha}[\cos(2 k_F x - \sqrt{2} (\varphi_\rho+\varphi_\sigma))
-\cos(2 k_F x - \sqrt{2} (\varphi_\rho-\varphi_\sigma))]
\right)=
\\
\frac{J_K}{4} \int dx
\frac{\sqrt{2} M^z_f \partial_x \varphi_\sigma}{\pi}
+
\frac{\sqrt{2}}{\pi^2}\partial_x\varphi(x)\partial_x \varphi_\sigma
\\
+
\frac{1}{(\pi\tilde\alpha)^2}\cos(2\varphi(x)-2k_F^f x)
[\cos(2 k_F x - \sqrt{2} (\varphi_\rho+\varphi_\sigma))
-\cos(2 k_F x - \sqrt{2} (\varphi_\rho-\varphi_\sigma))].
\end{gather*}

The first term is relevant and induces a magnetization in the itinerant subsystem leading to splitting of the two cases above into four $k_F^f =k_F\pm \frac{J_K M^z_f}{4 v_F},\;\pi-k_F\pm \frac{J_K M^z_f}{4 v_F}$. In the bosonic language this effect is absorbed into  $\varphi_\sigma\to \varphi_\sigma +\frac{J_K M^z_f}{\sqrt{2}v_F}x$.
On the other hand, the second term is marginal. As for the third term, let us first consider it on its own (without the effect of Zeeman splitting). This term is then nonvanishing only for $k_F^f = k_F;\;\pi-k_F$. In that case it can, however, be relevant:
\[
[e^{i2\varphi(x)}e^{i\sqrt{2}\varphi_\rho(x)}e^{i\sqrt{2}\varphi_\sigma(x)}] =
L^{-K-K_\rho/2-K_\sigma/2}.
\]
It is relevant if $K<1$ which is true throughout the superfluid phase. Consequently, this coupling is relevant. Rewriting the term for $k_F^f = k_F$ one has:
\[
H_{K,k_F}^{ZZ}=\frac{J_K}{4}\left[\frac{1/2}{(\pi\tilde\alpha)^2} \cos(2\varphi(x)-\sqrt{2}(\varphi_\rho(x)+\varphi_\sigma(x)))
-\frac{1/2}{(\pi\tilde\alpha)^2} \cos(2\varphi(x)-\sqrt{2}(\varphi_\rho(x)-\varphi_\sigma(x))),
\right]
\]
while for $k_F^f = \pi - k_F$:
\[
H_{K,\pi-k_F}^{ZZ}=\frac{J_K}{4}\left[\frac{1/2}{(\pi\tilde\alpha)^2}
\cos(2\varphi(x)+\sqrt{2}(\varphi_\rho(x)+\varphi_\sigma(x)))
-\frac{1/2}{(\pi\tilde\alpha)^2} \cos(2\varphi(x)+\sqrt{2}(\varphi_\rho(x)-\varphi_\sigma(x))).
\right]
\]
One can see that it acts to gap out two of the modes at filling where the itinerant fermions are commensurate with the excitations of localized spins. Note however that the Zeeman splitting itself is of order $J_K$. If we ignore it and go to the $k_F^f=k_F=\pi/2$ case there will be four terms probably gapping out all the modes in the system (see also analysis of this special case in \ref{sec:special}).

Taking into account the presence of both the Zeeman splitting and the cosine-interaction terms one sees that only one cosine term can be non-oscillating for a particular $h$, i.e. the following four cases $k_F^f =k_F\pm \frac{J_K M^z_f}{4 v_F},\;\pi-k_F\pm \frac{J_K M^z_f}{4 v_F}$, as is marked by the green regions in Fig. 1 (c) of the main text (the finite width in $h$ is there due to finite $J_K$).
Thus, to find out whether there are either four or two gapped phases we need to compare the Zeeman splitting with the gap due to the cosine term. If the Zeeman splitting is much smaller then the gap, one of the oscillating cosine terms (the one that oscillates like $\cos \frac{J_K M^z_f}{2 v_F} x$) can still open a gap similar to the doping-induced Mott transition\cite{giamarchi.2003}.

An estimate for the gap can be obtained with scaling arguments. The coupling constant of the cosine term grows as $e^{(1-K)l}$, $l$ being the RG scale, while the gap, having the dimensions of energy (scaling dimension $1$), as $e^{l}$. The RG flow is cut off when the dimensionless coupling becomes of order $1$ and the gap of the order of bandwidth, which results in the bare gap being of the order $\frac{v_F}{\tilde\alpha} \left(\frac{J_K}{v_F/\tilde\alpha}\right)^{\frac{1}{1-K}}$ that is parametrically smaller then $J_K$ until $J_K$ becomes of the order of the bandwidth. Consequently, the gap resulting from the cosine term is much weaker then the Zeeman splitting and there are indeed four separate special values in the phase diagram of Fig. 1 (c) (green regions) of the main text (the finite thickness there is due to finite $J_K$).

Away from the special fillings described above the $\langle \tilde{S}_z \tilde{S}_z\rangle$ correlations at $2k_F$ are short-ranged and one can use perturbation theory (note that the power-law correlations at zero momentum transfer lead to the gradient terms and are presented above). In second order one obtains the following term:
\begin{equation}
S_{ind}^{h>h_c}=-\frac{J_K^2}{32} \int dx \int d \tau\int dx' \int d \tau' \langle \overline{S}_z(x,\tau) \overline{S}_z(x',\tau')\rangle  [\psi^\dagger(x)\sigma_z \psi(x)][\psi^\dagger(x')\sigma_z \psi(x')],
\label{sup:eq:1bandintz}
\end{equation}
where $\overline{S}_z(x,\tau) = \tilde{S}_z(x,\tau)-\langle\tilde{S}_z(x,\tau)\rangle$ and $\langle \overline{S}_z(x,\tau) \overline{S}_z(x',\tau')\rangle$ has short-range correlations (however, unlike the $h<h_c$ case they are not zero here). Note that for finite $q$ one has $\int d(x-x') e^{iq [x-x']}\langle \overline{S}_z(x,\tau) \overline{S}_z(x',\tau')\rangle = \int d(x-x') e^{iq [x-x']} \langle \tilde{S}_z(x,\tau) \tilde{S}_z(x',\tau')\rangle$ as $\langle\tilde{S}_z(x)\rangle = const$  and thus the average terms do not contribute. For weak coupling this term is marginal. Assuming the interaction to be also instantaneous (which is a reasonable assumption if the spin correlations are gapped at $q$), one can rewrite it using the conventional notations\cite{giamarchi.2003} we get $g_{2,4\perp}=-g_{2,4\parallel}\equiv g >0$ and $g_{1\perp}=-g_{1\parallel}\equiv g' <g$ (as the momentum transfer for this term is $2k_F$ and $\langle S_z(x) S_z(x')\rangle_{q=2k_F}<\langle S_z(x) S_z(x')\rangle_{q=0}$) leading to $K_\rho=\left(\frac{1-g'/2}{1+g'/2}\right)<1$ and $K_\sigma=\left(\frac{1-g'/2}{1-2g+g'/2}\right)>1$, which leads to the conclusion that the interaction is marginally irrelevant. Note that for the case of half-filled band an umklapp term is allowed in the low-energy theory. As $K_\rho<1$ it follows that it is (marginally) relevant and leads to a Mott transition. In principle, for any commensurate filling an umklapp term will emerge in a certain order of perturbation theory. However, the requirement for Mott transition is $K_\rho<1/n^2$, $n$ being the order of commensurability. Thus, in the weak coupling limit the Mott transition is expected for the half filled band only.

The interband part of the Kondo Hamiltonian (XX and YY coupling), taken at low energies, results in the same ZZ coupling after the second band is integrated out, as is shown above, and does not lead to any new effects. The self-energy effects \eqref{eq:selfenZ} (where the large frequencies have to be taken into account) that lead to a Zeeman splitting remain present, however, their effect is perturbatively smaller in the most of the AFM phase than the one caused by the ZZ interaction with the spinons, the latter being linear in $J_K$.

\section{Two conduction bands at the Fermi Level}
\subsection{\texorpdfstring{$h<h_c$}{}}
The perturbative approach employed in the previous section does not work here, as the lowest-order diagram contains a logarithmically diverging fermionic bubble. On the other hand, the bosons are gapped in this regime and we can integrate them out to obtain an effective interaction for fermions. The low-energy theory for the conduction electrons (Fig. \ref{fig:2band}) is:
\begin{figure}[h!]
	\centering
	\includegraphics[width=0.8\linewidth]{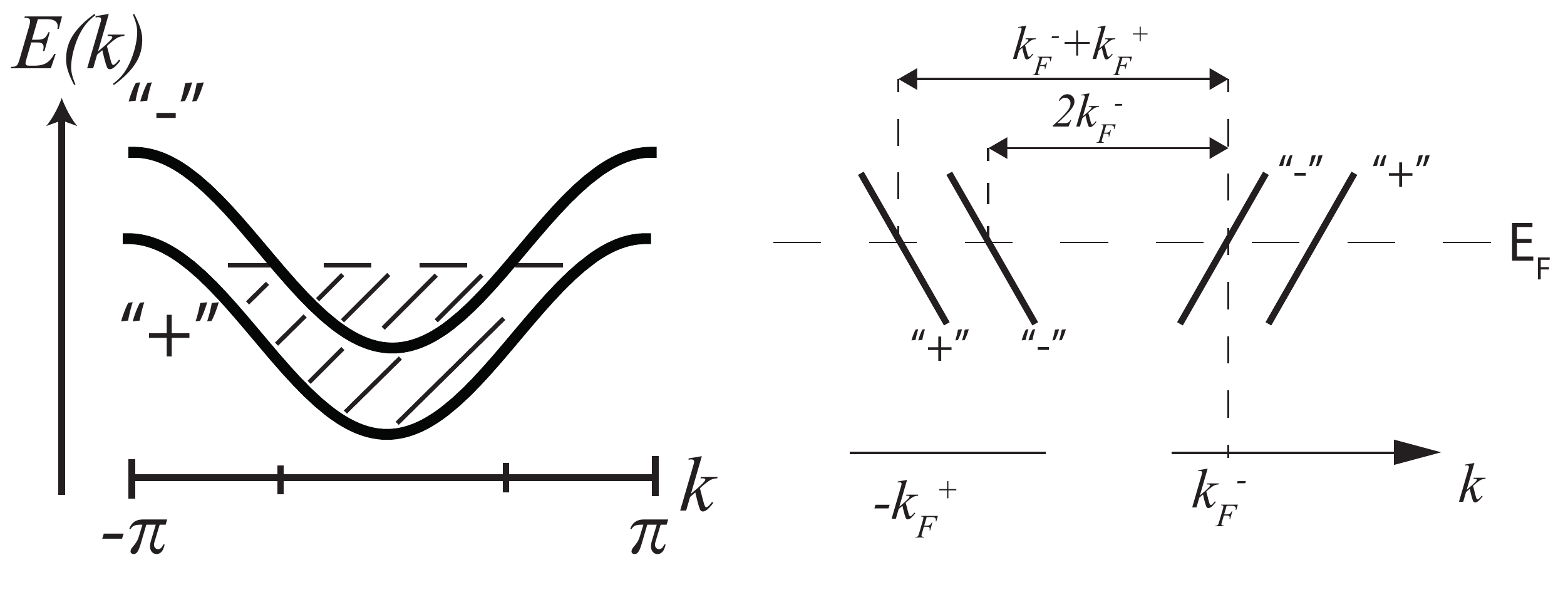}
	\caption{Linearization of dispersion for the conduction electrons.}
	\label{fig:2band}
\end{figure}

\[
H_0^\pm= \frac{1}{2\pi} \sum_{\alpha=\sigma,\rho} \int dx v_F^\pm [(\nabla \theta_\alpha^\pm(x))^2+(\nabla \varphi_\alpha^\pm(x))^2].
\]

First of all, the $S_z\sigma_z$ term in the Kondo Hamiltonian could generate an interaction for the fermions in second order in $J_K$. However, as was shown in the one-band case, the lowest-order contribution vanishes.

To integrate out the bosons with the remaining $S_+\sigma_-$ part of $H_K$ we simply perform an algebraic transformation in the action:
\begin{gather*}
S_{bos}+S_K^{XY} = (-i\omega + \varepsilon_q) a^\dagger_{q,\omega} a_{q,\omega} + a^\dagger_{q,\omega} j_{q,\omega} + a_{q,\omega} j^\dagger_{q,\omega} = (-i\omega + \varepsilon_q) \tilde{a}^\dagger_{q,\omega} \tilde{a}_{q,\omega}
-(-i\omega + \varepsilon_q)^{-1}j^\dagger_{q,\omega}j_{q,\omega};
\\
\tilde{a}_{q,\omega} = a_{q,\omega}+(i\omega - \varepsilon_q)^{-1}j_{q,\omega};
\\
j_{q,\omega} = \frac{J_K}{2\sqrt{2}}
\frac{1}{\sqrt{N}}\sum_k
\left(
\psi_{+,\downarrow}^\dagger(k-q,\varepsilon-\omega)\psi_{-,\uparrow}(k,\varepsilon)
+
\psi_{-,\downarrow}^\dagger(k-q,\varepsilon-\omega)\psi_{+,\uparrow}(k,\varepsilon)
\right).
\end{gather*}
The interactions lead to two important effects : a) appearance of self-energies that ultimately results in a Zeeman effect for the conduction electrons; b) corrections to the interactions contain low-energy singularities and may result in gap opening, if some of them flow to strong coupling in the RG sense. a) has been considered in \ref{sec:selfen} and the expressions derived there (e.g. for the Zeeman splitting Eq. \eqref{eq:selfenZ}) remain valid also in the two band case.

\subsection{Low-energy interactions and instabilities}
Here we ignore the self-energy effects discussed in \ref{sec:selfen} and concentrate on the low-energy interaction between the fermions of the two bands.
As the bosonic spectrum $\varepsilon_q$ is gapped, to extract the low-energy lagrangian we can take $\omega=0$ and $q$ such that all the fermionic fields are close to the Fermi momentum in the inverse propagator $(-i\omega+ \varepsilon_q)^{-1}$ (the neglected terms will have higher powers of $q$ or $\omega$ making them less relevant). As there is no $\omega$ dependence we can write an explicit interaction Hamiltonian. The possible $q$ values are $\pm (k_F^++k_F^-),\pm(k_F^+-k_F^-)$. The resulting Hamiltonian is (using the usual 1D notations for right- and left- movers):

\begin{gather*}
S_{int} = -\sum_{k,k',q}
\frac{J_K^2}{8N\varepsilon_{q}}
\left(
\psi_{+,\uparrow}^\dagger(k+q,\varepsilon-\omega)\psi_{-,\downarrow}(k,\varepsilon)
+
\psi_{-,\uparrow}^\dagger(k+q,\varepsilon-\omega)\psi_{+,\downarrow}(k,\varepsilon)
\right)
\\
\times
\left(
\psi_{+,\downarrow}^\dagger(k'-q,\varepsilon-\omega)\psi_{-,\uparrow}(k',\varepsilon)
+
\psi_{-,\downarrow}^\dagger(k'-q,\varepsilon-\omega)\psi_{+,\uparrow}(k',\varepsilon)
\right)\approx
\\
-\frac{J_K^2}{8N\varepsilon_{k_F^1-k_F^2}}\sum_{\tilde{k},\tilde{k}',\tilde{q},r=R,L}
\left[
\psi_{+,\uparrow}^{r,\dagger}(k+q,\varepsilon-\omega)\psi^r_{-,\downarrow}(k,\varepsilon)
\psi_{-,\downarrow}^{r,\dagger}(k'-q,\varepsilon-\omega)\psi^{r}_{+,\uparrow}(k',\varepsilon)
\right.
\\
+
\psi_{-,\uparrow}^{r,\dagger}(k+q,\varepsilon-\omega)\psi^r_{+,\downarrow}(k,\varepsilon)
\psi_{+,\downarrow}^{r,\dagger}(k'-q,\varepsilon-\omega)\psi^{r}_{-,\uparrow}(k',\varepsilon)
\\
+\psi^{\dagger,r}_{+,\uparrow}(k+q,\varepsilon-\omega)\psi^r_{-,\downarrow}(k,\varepsilon)
\psi^{\dagger,-r}_{+,\downarrow}(k'-q,\varepsilon-\omega)\psi^{-r}_{-,\uparrow}(k',\varepsilon)
\\
\left.
\psi_{-,\uparrow}^{r,\dagger}(k+q,\varepsilon-\omega)\psi^r_{+,\downarrow}(k,\varepsilon)
\psi_{-,\downarrow}^{-r,\dagger}(k'-q,\varepsilon-\omega)\psi^{-r}_{+,\uparrow}(k',\varepsilon)
\right]
\\
-\frac{J_K^2}{8N\varepsilon_{k_F^1+k_F^2}}\sum_{\tilde{k},\tilde{k}',\tilde{q},r=R,L}
\left[
\psi_{+,\uparrow}^{r,\dagger}(k+q,\varepsilon-\omega)\psi^{-r}_{-,\downarrow}(k,\varepsilon)
\psi_{+,\downarrow}^{-r,\dagger}(k'-q,\varepsilon-\omega)\psi^r_{-,\uparrow}(k',\varepsilon)
+
\right.
\\
+
\psi_{-,\uparrow}^{r,\dagger}(k+q,\varepsilon-\omega)\psi^{-r}_{+,\downarrow}(k,\varepsilon)
\psi_{-,\downarrow}^{-r,\dagger}(k'-q,\varepsilon-\omega)\psi^r_{+,\uparrow}(k',\varepsilon)
\\
+
\psi_{+,\uparrow}^{r,\dagger}(k+q,\varepsilon-\omega)\psi^{-r}_{-,\downarrow}(k,\varepsilon)
\psi_{-,\downarrow}^{-r,\dagger}(k'-q,\varepsilon-\omega)\psi^r_{+,\uparrow}(k',\varepsilon)
\\
+
\left.
\psi_{-,\uparrow}^{r,\dagger}(k+q,\varepsilon-\omega)\psi^{-r}_{+,\downarrow}(k,\varepsilon)
\psi_{+,\downarrow}^{-r,\dagger}(k'-q,\varepsilon-\omega)\psi^r_{-,\uparrow}(k',\varepsilon)
\right],
\end{gather*}
where $\tilde{k},\tilde{k}',\tilde{q}$ means the sum being taken only over small momenta. Rewriting these in real space one gets:
\begin{equation}
\begin{gathered}
S_{int}=\int dx
\left[
\frac{J_K^2}{8\varepsilon_{k_F^1-k_F^2}}
\sum_{r=L,R}
\rho^r_{+,\uparrow}(x)\rho^r_{-,\downarrow}(x)+\rho^r_{+,\downarrow}(x)\rho^r_{-,\uparrow}(x)
+
\left(
\psi_{+,\uparrow}^{r,\dagger}(x)\psi_{+,\downarrow}^{-r,\dagger}(x)\psi^{r}_{-,\downarrow}(x)
\psi^{-r}_{-,\uparrow}(x)+h.c.
\right)
\right]
\\
+
\left[
\frac{J_K^2}{8\varepsilon_{k_F^1+k_F^2}}
\sum_{r=L,R}
\rho^{-r}_{+,\uparrow}(x)\rho^r_{-,\downarrow}(x)+\rho^{-r}_{+,\downarrow}(x)\rho^r_{-,\uparrow}(x)
+
\left(
\psi_{+,\uparrow}^{r,\dagger}(x)\psi_{+,\downarrow}^{-r,\dagger}(x)\psi^{-r}_{-,\downarrow}(x)
\psi^r_{-,\uparrow}(x)+h.c.
\right)
\right]
.
\end{gathered}
\label{eq:2bandint}
\end{equation}
The expression above can be bosonized yielding:
\begin{gather*}
S_{int} = \int dx
\frac{J_K^2}{8\varepsilon_{k_F^1-k_F^2}}
\frac{1}{\pi^2}
\left[
\nabla \varphi_\rho^+\nabla \varphi_\rho^-
-
\nabla \varphi_\sigma^+\nabla \varphi_\sigma^-
\right]
\\
+
\frac{J_K^2}{8\varepsilon_{k_F^+ +k_F^-}}
\frac{1}{2\pi^2}
\left[
\nabla \varphi_\rho^+\nabla \varphi_\rho^-
-
\nabla \varphi_\sigma^+\nabla \varphi_\sigma^-
-
\nabla \theta_\rho^+\nabla \theta_\rho^-
+
\nabla \theta_\sigma^+\nabla \theta_\sigma^-
\right]
\\
+
\frac{J_K^2}{8\varepsilon_{k_F^+-k_F^-}\pi^2\tilde\alpha^2}
\cos \sqrt{2}(\theta^+_\rho - \theta^-_\rho)
\cos \sqrt{2} (\varphi_\sigma^+ +\varphi_\sigma^-)
\\
+
\frac{J_K^2}{8\varepsilon_{k_F^+ +k_F^-}\pi^2\tilde\alpha^2}
\cos \sqrt{2}(\theta^+_\rho - \theta^-_\rho)
\cos \sqrt{2} (\varphi_\sigma^+ -\varphi_\sigma^-).
\end{gather*}

The first two terms contribute to the renormalization of the Fermi velocities as well as Luttinger parameter $K$, which are marginal in the RG sense while the last two are genuine interaction terms. Their scaling dimension is:
\[
[\cos \sqrt{2}(\theta^+_\rho - \theta^-_\rho)
\cos \sqrt{2} \varphi_\sigma^+ \cos \sqrt{2} \varphi_\sigma^-]=
L^{-\left(\frac{1}{2 K_\rho^+}+\frac{1}{2 K_\rho^-}+\frac{K_\sigma^+}{2}+\frac{ K_\sigma^-}{2}\right)}.
\]
It is evident that in the absence of other interactions these terms are also marginal. Thus to move forward we should consider the RG flow with the initial conditions specified by (\ref{eq:2bandint}). The 2-loop RG equations for an interacting 2-band model have been derived for the general case in \cite{varma.1985,penc.1990}. Note that the more compact form used in \cite{balents.1996} is derived using SU(2) spin symmetry not present in our case and consequently we have to use the more complicated general form. Of all the relevant couplings considered there, in the current model we have only $g_{ABBA}^{\perp(2)} = \frac{J_K^2}{8\varepsilon_{k_F^++k_F^-}},\;g_{AABB}^{\perp(2)}= \frac{J_K^2}{8\varepsilon_{k_F^++k_F^-}},\;g_{AABB}^{\perp(1)}= \frac{J_K^2}{8\varepsilon_{k_F^+-k_F^-}}$, where $A$ and $B$ are $+$ and $-$ in our case. However, other 9 couplings get generated in the RG. An example of an RG flow is in Fig. \ref{fig:RGrun}.

\begin{figure}[h!]
	\centering
	\includegraphics[width=0.8\linewidth]{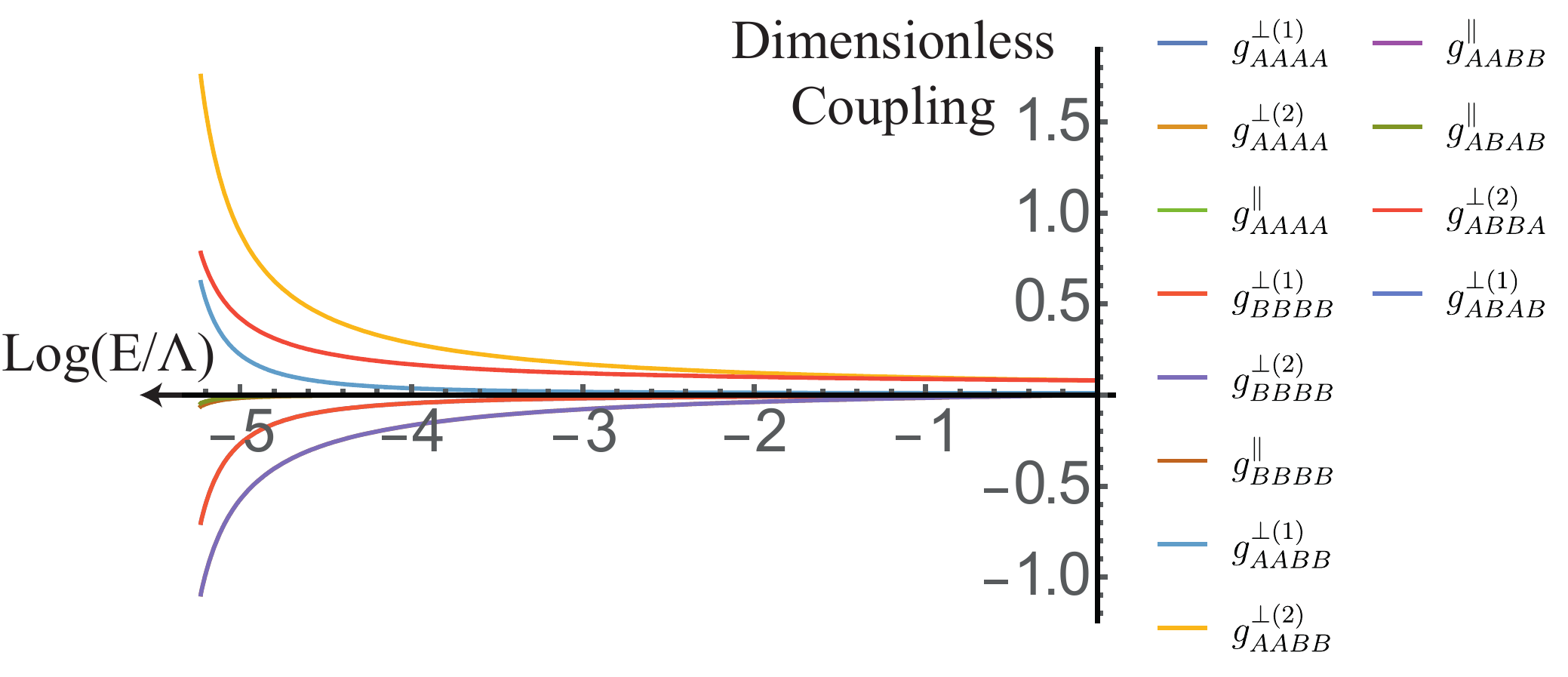}
	\caption{Running coupling constants as a function of the RG scale with initial dimensionless couplings $\frac{\tilde\alpha}{\pi\hbar (v_F^++v_F^-)} g_{ABBA}^{\perp(2)}= \frac{\tilde\alpha}{\pi\hbar (v_F^++v_F^-)} g_{AABB}^{\perp(2)}=0.08,\;\frac{\tilde\alpha}{\pi\hbar (v_F^++v_F^-)}g_{AABB}^{\perp(1)}=0.01$, $v_F^+/v_F^-=2$. The flow is shown up to $0.96$ of the value of RG scale where couplings diverge. Some of the couplings are equal and the corresponding curves overlap - see text for the description of the flow. Qualitatively similar results are obtained for other combinations of parameter values.}
	\label{fig:RGrun}
\end{figure}

It turns out that 10 of 12 couplings grow and eventually diverge at a critical scale. Two of these ($g_{AAAA}^{\perp(1)}=g_{BBBB}^{\perp(1)}$) correspond to the cosine term in usual 1D fermion chains with spin. They lead to the pinning of the phase $\varphi_\sigma$. Consequently, the spin correlations induced by the conduction electrons are short-ranged in this phase. However, to study the qualitative character of the resulting phase in detail we need to bosonize the interaction terms that diverge. It goes as follows:
\[
\psi_{r=R,L;\sigma}^\pm(x) =\frac{ U_{r,\sigma}^\pm}{\sqrt{2\pi \tilde\alpha}} e^{i r k_F x} e^{-\frac{i}{\sqrt{2}}[r \varphi_\rho(x)-\theta_\rho(x)+\sigma(r\varphi_\sigma(x)-\theta_\sigma(x))]^\pm},
\]
where $\tilde\alpha$ is the lattice cutoff scale. First we omit the Klein factors and write the bosonized expression corresponding to each of the interactions (we us the notations of \cite{penc.1990}).Below $\pm_r = +$ for $r=R$ and $-$ otherwise, $\pm_\sigma = +$ for $\sigma=\uparrow$ and $-$ otherwise; we also omit the band indices $\pm$ where all the operators belong to the same band:
\begin{gather*}
\rho_{r,\sigma}^\pm = \frac{1}{2\sqrt{2}\pi} \nabla\left(
\pm_r[\theta_\rho^r\pm_\sigma\theta_\sigma^r] - [\varphi_\rho^r\pm_\sigma \varphi_\sigma^r]
\right),
\\
g^{\perp(1)}_{AAAA}:
\left[\sum_{r,\sigma}
:\psi_{r,\sigma}^{\dagger}\psi_{-r,\sigma} \psi_{-r,-\sigma}^\dagger\psi_{r,-\sigma}:
\right]^\pm
=
\frac{1}{(2\pi\tilde\alpha)^2} \sum_{r,\sigma,a}
e^{\frac{i}{\sqrt{2}}[-2\theta_\rho(x)+2\sigma r\varphi_\sigma^a(x)]}
e^{-\frac{i}{\sqrt{2}}[-2\theta_\rho(x)-2\sigma r\varphi_\sigma^{-a}(x)]}
\\
=
\frac{1}{(\pi\tilde\alpha)^2}
\cos(2\sqrt{2}\varphi_\sigma(x));
\\
g^{\perp(2)}_{AAAA}:
\left[\sum_{r,\sigma} \psi_{r,\sigma}^\dagger\psi_{r,\sigma} \psi_{-r,-\sigma}^\dagger\psi_{-r,-\sigma}\right]^\pm
=
\frac{1}{2\pi^2}
\left[(\nabla \varphi_\rho)^2-(\nabla \varphi_\sigma)^2+(\nabla \theta_\sigma)^2-(\nabla \theta_\rho)^2\right]^\pm;
\\
g^\parallel_{AAAA}:
\left[\sum_{r,\sigma} :\psi_{r,\sigma}^\dagger\psi_{-r,\sigma} \psi_{-r,\sigma}^\dagger\psi_{r,\sigma}:\right]^\pm
=
-\left[\sum_{r,\sigma} :\psi_{r,\sigma}^\dagger\psi_{r,\sigma}\psi_{-r,\sigma}^\dagger \psi_{-r,\sigma}: \right]^\pm
=
\\
=-\frac{1}{2\pi^2}
\left[(\nabla \varphi_\rho)^2+(\nabla \varphi_\sigma)^2-(\nabla \theta_\sigma)^2-(\nabla \theta_\rho)^2\right]^\pm;
\end{gather*}

\begin{gather*}
g^{\perp(1)}_{AABB}:
\left[\sum_{r,\sigma,a=+,-}
:\psi_{r,\sigma}^{a,\dagger}\psi_{-r,\sigma}^{-a} \psi_{-r,-\sigma}^{a,\dagger}\psi_{r,-\sigma}^{-a}:
\right]
=
\frac{1}{(2\pi\tilde\alpha)^2} \sum_{r,\sigma,a}
e^{\frac{i}{\sqrt{2}}[-2\theta_\rho^a(x)+2\sigma r\varphi_\sigma^a(x)]}
e^{-\frac{i}{\sqrt{2}}[-2\theta_\rho^{-a}(x)-2\sigma r\varphi_\sigma^{-a}(x)]}
\\
=
\frac{2}{(\pi\tilde\alpha)^2}
\cos(\sqrt{2}[\theta_\rho^+(x)-\theta_\rho^-(x)])
\cos(\sqrt{2}[\varphi_\sigma^+(x)+\varphi_\sigma^-(x)]);
\\
g^{\perp(2)}_{AABB}:
\left[\sum_{r,\sigma,a=+,-}
:\psi_{r,\sigma}^{a,\dagger}\psi_{r,\sigma}^{-a} \psi_{-r,-\sigma}^{a,\dagger}\psi_{-r,-\sigma}^{-a}:
\right]
=
\frac{1}{(2\pi\tilde\alpha)^2} \sum_{r,\sigma,a}
e^{\frac{i}{\sqrt{2}}[-2\theta_\rho^a(x)+2\sigma r\varphi_\sigma^a(x)]}
e^{-\frac{i}{\sqrt{2}}[-2\theta_\rho^{-a}(x)+2\sigma r\varphi_\sigma^{-a}(x)]}
\\
=
\frac{2}{(\pi\tilde\alpha)^2}
\cos(\sqrt{2}[\theta_\rho^+(x)-\theta_\rho^-(x)])
\cos(\sqrt{2}[\varphi_\sigma^+(x)-\varphi_\sigma^-(x)]);
\\
g^{\parallel}_{ABAB}:
\left[\sum_{r,\sigma,a=+,-}
:\psi_{r,\sigma}^{a,\dagger}\psi_{-r,\sigma}^{-a} \psi_{-r,\sigma}^{-a,\dagger} \psi_{r,\sigma}^{a}:
\right]
=
-\left[\sum_{r,\sigma,a=+,-}
:\psi_{r,\sigma}^{a,\dagger}\psi_{r,\sigma}^{a} \psi_{-r,\sigma}^{-a,\dagger}\psi_{-r,\sigma}^{-a}: \right]
=
\\
=-\frac{1}{\pi^2}
\left[\nabla \varphi_\rho^+\nabla \varphi_\rho^- + \nabla \varphi_\sigma^+\nabla \varphi_\sigma^-
-\nabla \theta_\sigma^+\nabla \theta_\sigma^- -\nabla \theta_\rho^+\nabla \theta_\rho^-\right];
\\
g^{\perp(2)}_{ABBA}:
\left[\sum_{r,\sigma,a=+,-}
:\psi_{r,\sigma}^{a,\dagger}\psi_{r,\sigma}^{a} \psi_{-r,-\sigma}^{-a,\dagger} \psi_{-r,-\sigma}^{-a}:
\right]
=
\frac{1}{\pi^2}
\left[\nabla \varphi_\rho^+\nabla \varphi_\rho^- - \nabla \varphi_\sigma^+\nabla \varphi_\sigma^-
+\nabla \theta_\sigma^+\nabla \theta_\sigma^- -\nabla \theta_\rho^+\nabla \theta_\rho^-\right],
\end{gather*}
where $:...:$ is for normal ordering; each term is marked by the g-notation from \cite{penc.1990} on the LHS; on the RHS there is an expression in terms of fermionic operators and an equivalent bosonized form (without the Klein factors).

Let us now consider the effects of the presence of Klein factors $U^a_{r,\sigma}$ in the operators $\psi_{r,\sigma}^a$. The $U$ operators can be chosen unitary such that for those with the same indices they satisfy $U^{a\dagger}_{r,\sigma}U^{a}_{r,\sigma}=U^{a}_{r,\sigma}U^{a\dagger}_{r,\sigma}=1$\cite{haldane.1981}. Operators with different indices, on the other hand, anticommute $(U^{a}_{r,\sigma})'U^{a\dagger}_{r,\sigma}=-U^{a\dagger}_{r,\sigma}(U^{a}_{r,\sigma})'$.
For these factors to be safely omitted in the interaction term it should be possible to diagonalize them in all the interaction terms simultaneously. Then, as the eigenvalues of these terms can be shown (most easily using Majorana representation for $U$) to be $\pm1$, the products of Klein factors in the interaction terms reduce to just a sign\cite{schulz.1996}. The sign can be chosen arbitrarily, but should be consistent with signs chosen for the observables\cite{giamarchi.2003}.

Let us first consider the issue of simultaneous diagonalization. Actually, for many of the couplings above the products of Klein factors reduce to $\pm1$ due to $U^\dagger U=1$ (the ones that have $-1$ due to the interchange of fermionic operators are highlighted above). The ones remaining are: $g^{\perp(1)}_{AAAA},\;g^{\perp(1)}_{AABB},\;g^{\perp(2)}_{AABB}$. Every pair of them has only two operators with the same indices, implying that these terms commute with each other\cite{schulz.1996}.

Let us now consider the signs. More precisely, we need to define the observables in agreement with the sign conventions for the interactions. We choose the Klein factor products in the interactions to be all equal to $1$. Consequently, the CDW/SDW/SC operators in a single leg are defined as for the single fermionic chain in\cite{giamarchi.2003}:
\begin{equation}
\begin{gathered}
O_{CDW}^\pm(x)=\frac{e^{-2ik_F^\pm x}}{\pi\tilde\alpha}e^{i\sqrt{2}\varphi_\rho^\pm}\cos(\sqrt{2}\varphi_\sigma^\pm);
\\
O_{SDWz}^\pm(x)=\frac{e^{-2ik_F^\pm x}}{\pi \tilde\alpha}e^{i\sqrt{2}\varphi_\rho^\pm}i\sin(\sqrt{2}\varphi_\sigma^\pm);
\\
O_{SC}^\pm =\frac{1}{\pi\tilde\alpha}e^{-i\sqrt{2}\theta_\rho^\pm}\cos(\sqrt{2}\varphi_\sigma^\pm).
\end{gathered}
\label{eq:observ}
\end{equation}
Regarding the possible phases there are two possible tuning parameters: ratios $\varepsilon_{k_F^++k_F^-}/\varepsilon_{k_F^+-k_F^-}$ and $v_F^+/v_F^-$ (the absolute value of the bare couplings should not matter as long as they are small). The system always flows to strong coupling and eventually diverges at some critical scale $l_0$. The signs of the couplings are set well before the coupling become of order 1 and do not change up to $l_c$. The signs are: $g^{\perp(1)}_{++++}=g^{\perp(1)}_{----}<0;\;g^{\perp(2)}_{++++}=g^{\perp(2)}_{----}<0;
\;g^{\parallel}_{++++}=g^{\parallel}_{----}<0;\;g^{\perp(1)}_{++--}>0;\;g^{\perp(2)}_{++--}>0;\;
g^{\parallel}_{+-+-}<0;\;g^{\perp(2)}_{+--+}<0$. Of these couplings $g^{\parallel}_{++++},g^{\parallel}_{+-+-}<0,g^{\perp(2)}_{+--+}$ turn out to be smaller in absolute value then the others.

Using the bosonized expressions above one finds that $g^{\perp(1)}_{++++/----}<0$ pins the $\varphi_\sigma^\pm$ fields and then the $g^{\perp(1)}_{++--}>0,\;g^{\perp(2)}_{++--}>0$ fix $\theta_\rho^+ - \theta_\rho^-$, leaving only the total charge mode $\theta_\rho^+ + \theta_\rho^-$ ungapped. This fixes $\varphi_{\sigma}^\pm$ to $n_\pm*\pi/\sqrt{2}$ and $\sqrt{2}[\theta_\rho^+(x)-\theta_\rho^-(x)]$ to $2n\pi$ if $n_+-n_-$ is odd and $(2n+1)\pi$ if $n_+-n_-$ is even. The expressions for observables (\ref{eq:observ}) immediately yield that there are CDW and SC correlations. Moreover, the values of $\theta_\rho^+(x)-\theta_\rho^-(x)$ and $\varphi_{\sigma}^\pm$ dictate that $O_{SC}^+ = -O_{SC}^-$, showing the d-wave character of the state. It is straightforward to check that these results are in line with a naive mean-field decoupling of the interaction terms, signifying that we have chosen consistent signs due to the Klein factors. To be even more precise, let us consider the CDW-SC competition controlled by the Luttinger parameter $K_\rho$. Numerical solutions of RG equation that the $g^{\perp(2)}_{AAAA/BBBB}<0$ couplings are by far dominant in absolute value with regard to other coupling resulting in gradient terms. This leads to $K_\rho>1$, i.e. SC correlations $\sim r^{-1/2K_\rho}$ are dominant. Also, $K_\sigma$ becomes less then $1$ in the RG flow consistent with the single chain result\cite{giamarchi.2003}.

Let us now discuss the interplay of the above effects and the Zeeman splitting arising from self-energy in Eq. \eqref{eq:selfenZ}. Zeeman splitting actually hinders superconductivity (see also below); in the RG sense, it provides a lower cutoff for the flow. At that cutoff, if the dimensionless couplings responsible for the superconductivity are still small, the Zeeman effect needs to be taken into account first - it results in the absence of superconductivity in the low-energy theory (see below). However, if the coupling become of order 1 at the cutoff scale this suggests that superconducting corrrelations overcome the Zeeman splitting and  dominate the low-energy physics. The couplings are marginal such that $g(l)\sim g_0 l$ where the RG scale $l$ is equal to $\log (v_F/\tilde\alpha)/E_Z$ at the cutoff, resulting in the condition $v_F/\tilde\alpha e^{-1/g_0}> E_Z$ for a Zeeman energy $E_Z$. Same condition can be obtained by comparing an estimate for the superconducting gap with the Zeeman energy akin to the Clogston-Chandrasekar limit \cite{clogston.1962,chandrasekar.1962}. Note that in our case both $E_Z$ and $g_0$ are quadratic in $J_K$; this suggests that for infinitesimal $J_K$ the Zeeman effect dominates, while at larger $J_K$ the spin gap due to superconductivity may take over.

\subsection{\texorpdfstring{$h>h_c$}{}}
Apart from the special cases $k_F^+, k_F^-=\pi,0,\pi\pm2\pi m,\pm2\pi m$ there is no interaction between the low-energy fermions and spinons due to the XX and YY parts of the Kondo coupling. As the spin correlations at $q=\pm (k_F^+ \pm k_F^-)$ are gapped one could expect similar physics as for $h<h_c$ albeit with a Zeeman splitting of $J_K M^z_f/2$. Like in the single band case, the Zeeman splitting has to be included before other interaction effects, its scaling dimension being larger.

The Zeeman splitting reduces the number of low-energy couplings for the model respecting momentum conservation, as now $k_F^{\pm\uparrow}\neq k_F^{\pm\downarrow}$. Namely, $g_{AAAA/BBBB}^{\perp(1)},\;g_{ABAB}^{\perp(1)},\;g_{AABB}^{\perp(1)},\;g_{AABB}^{\perp(2)}$ involve at least one fermionic operator not at the Fermi energy. The latter two are actually induced by the Kondo coupling and correspond to the cosine terms in bosonized form. Thus the only low-energy interaction remaining is $g^{\perp(2)}_{ABBA}$ that renormalizes the velocities but can not open a gap.

One can also analyze the special points $k_F^f = k_F^\pm;\; \pi - k_F^\pm$ or $k_F^f = k_F^- \pm k_F^+;\; \pi -(k_F^- \pm k_F^+)$. The first case is identical to the one for the single chain, i.e. there appears an interaction term (ZZ) that is relevant. For the case $k_F^f =  k_F^- + k_F^+$ the interaction term appears (using (\ref{eq:spinboson})):
\begin{gather*}
H_K^{XY} = -\frac{J_K}{2\sqrt{2}} \sum_i\tilde{S}^-_i(\psi^\dagger_{+,i}\sigma^+\psi_{-,i}+\psi^\dagger_{-,i}\sigma^+\psi_{+,i})+
\tilde{S}^+_i(\psi^\dagger_{+,i}\sigma^-\psi_{-,i}+\psi^\dagger_{-,i}\sigma^-\psi_{+,i})
\\
\to
-\frac{J_K}{2\sqrt{2}} \int dx
\frac{e^{-i\theta(x)}}{\sqrt{2\pi\tilde\alpha}}\cos(2\varphi(x)-2(k_F^f+\pi/a)x)
\frac{1}{\sqrt{2\pi \tilde\alpha}}
e^{-i r k_F^- x}
e^{\frac{i}{\sqrt{2}}[r \varphi_\rho(x)-\theta_\rho(x)+\sigma(r\varphi_\sigma(x)-\theta_\sigma(x))]^-}
\\
\frac{1}{\sqrt{2\pi \tilde\alpha}} e^{-i r k_F^+ x} e^{-\frac{i}{\sqrt{2}}[-r \varphi_\rho(x)-\theta_\rho(x)+\sigma(-r\varphi_\sigma(x)-\theta_\sigma(x))]^+}
+c.c.
\end{gather*}
The expression above is needed only to evaluate the scaling dimension which is $2-\frac{1}{4K} - K - (K_\rho+K_\rho^{-1})/4-(K_\sigma+K_\sigma^{-1})/4<0$ for $3/4<K<1$, i.e. is irrelevant. Interestingly, the long-range spin correlations seem to suppress the instability of the fermion LL. Note that in this special case the spin correlations at $k_F^--k_F^+$ are gapped and can be still integrated out (vice versa for $k_F^f =  k_F^- - k_F^+$). Running the RG discussed above for $g^{\perp(1)}_{AAAA}=0,\;g^{\perp(2)}_{AABB}=0$ ($g^{\perp(1)}_{AABB}=0$) we get the results that $g^{\perp(2)}_{++++}=g^{\perp(2)}_{----}<0$ and $g^{\perp(2)}_{+--+}<0,\;g^{\perp(1)}_{++--}>0$ ($g^{\perp(2)}_{+--+}>0,\;g^{\perp(2)}_{++--}>0$) leading to 
a phase with one charge and one spin mode (although with a mixed character in band space) being gapped
($g^{\perp(2)}_{++--}$ gaps out the modes, while the other couplings only renormalize the LL parameters).

\section{Special fillings of the conduction bands}
\label{sec:special}
The following particular cases are special:

$\bullet$ {\it 1 band at the Fermi level $k_F=\pi/2$ (or other commensurate filling)}:

For $h>h_c$ there is a short-ranged ZZ coupling at $2 k_F$ generated by the localized spins away from $k_F^f=\pi/2 \leftrightarrow \tilde{h}=0$. One can consider now also the umklapp terms:
\[
S_{ind,umkl}^{h>h_c} = -\frac{J_K^2}{32} \int dx \langle S_z S_z\rangle  (\Psi^\dagger(x)\sigma_z \Psi(x))^2
\sim e^{\frac{i}{\sqrt{2}}(2 \varphi_\rho-2\theta_\rho)}
e^{-\frac{i}{\sqrt{2}}(-2 \varphi_\rho-2\theta_\rho)} +c.c.= \cos(2\sqrt{2} \varphi_\rho),
\]
with the scaling dimension $2 - 2 K_\rho$. As $K_\rho<1$ (see discussion below Eq. (\ref{sup:eq:1bandintz})) this term is relevant and leads to a Mott transition. The phase $\varphi_\rho$ is then pinned at $\pi/\sqrt{2}$ and the CDW as well as SC correlations are short-ranged while all the spin correlations are power-law and isotropic\cite{voit.1992}.

Other commensurate fillings could be analyzed similarly, but one needs to go to higher orders in perturbation theory in $J_K$ and thus we do not consider them. Moreover, Mott transition for non-half-filled cases are not expected to occur at arbitrary weak coupling as $K_\rho<1/n^2$, where $n$ is the order of commensurability, is required\cite{giamarchi.2003}, unlike the half-filled case.

$\bullet$ {\it 2 bands at the Fermi level with $k_F^++k_F^-=\pi$ (note that $k_F^+-k_F^-=\pi$ is not possible)}:

For $h<h_c$ we should additionally consider the umklapp interactions. This is however rather complicated, as this adds several new coupling constants to the RG and the equations in \cite{penc.1990} were derived for incommensurate filling. For the Hubbard model \cite{balents.1996} the umklapps favor the fully localized Mott state C0S0.

In this case the nature of the 'BEC' transition is not as clear, as the gap is indeed crucial for itinerant fermion correlations at $q=\pi$.

For $h>h_c$ one can note that in the bosonized expressions for $S^\pm(x)$ there is a part oscillating with $q=\pi$ throughout the superfluid phase. This results in an additional coupling:
\begin{gather*}
-\frac{J_K}{2\sqrt{2}} \int dx
\frac{e^{-i\theta(x)}}{(\sqrt{2\pi\tilde\alpha})^3}
\left(
\sum_{r,\sigma,a=\pm}
e^{\frac{i}{\sqrt{2}}[r \varphi_\rho(x)-\theta_\rho(x)+\sigma(r\varphi_\sigma(x)-\theta_\sigma(x))]^a}
e^{-\frac{i}{\sqrt{2}}[-r \varphi_\rho(x)-\theta_\rho(x)-\sigma(-r\varphi_\sigma(x)-\theta_\sigma(x))]^{(-a)}}
\right)
\\
+c.c.
\end{gather*}
The scaling dimension of this term is $2-\frac{1}{4K} - (K_\rho+K_\rho^{-1})/4-(K_\sigma+K_\sigma^{-1})/4>0$ for $3/4<K<1$ and small $J_K$. Thus this term is relevant. Performing the summations we get:
\[
\sim\cos[(\varphi_\rho^++\varphi_\rho^-+\varphi_\sigma^+-\varphi_\sigma^-)/\sqrt{2}]
\cos[(\theta_\sigma^++\theta_\sigma^-)/\sqrt{2}]
\cos[(\theta_\rho^+-\theta_\rho^-)/\sqrt{2}]\cos(\theta).
\]
Since the expression contains non-commmuting fields it cannot be interpreted in semiclassical terms. One sees that it can potentially localize four out of five gapless modes in the system.

$\bullet$ $k_F^+=0$:

In this case the second band is exactly at a Van Hove singularity. Consequently, one could think that the results of the treatment for $k_F^+\neq 0$ still apply, but with a certain enhancement of the interaction effects due to the singularity in the density of states.

\section{Extension to 2D/3D}
We will show now that the conclusion of the stability of the Magnon BEC transition (at least in the 1-band case) carries over to higher dimensionalities. Let us consider the relevance of the Kondo coupling for the case of a 2 or 3 dimensional system built of ladders (a 2D example id in Fig. \ref{fig:2d}).
\begin{figure}[h!]
	\centering
	\includegraphics[width=0.8\linewidth]{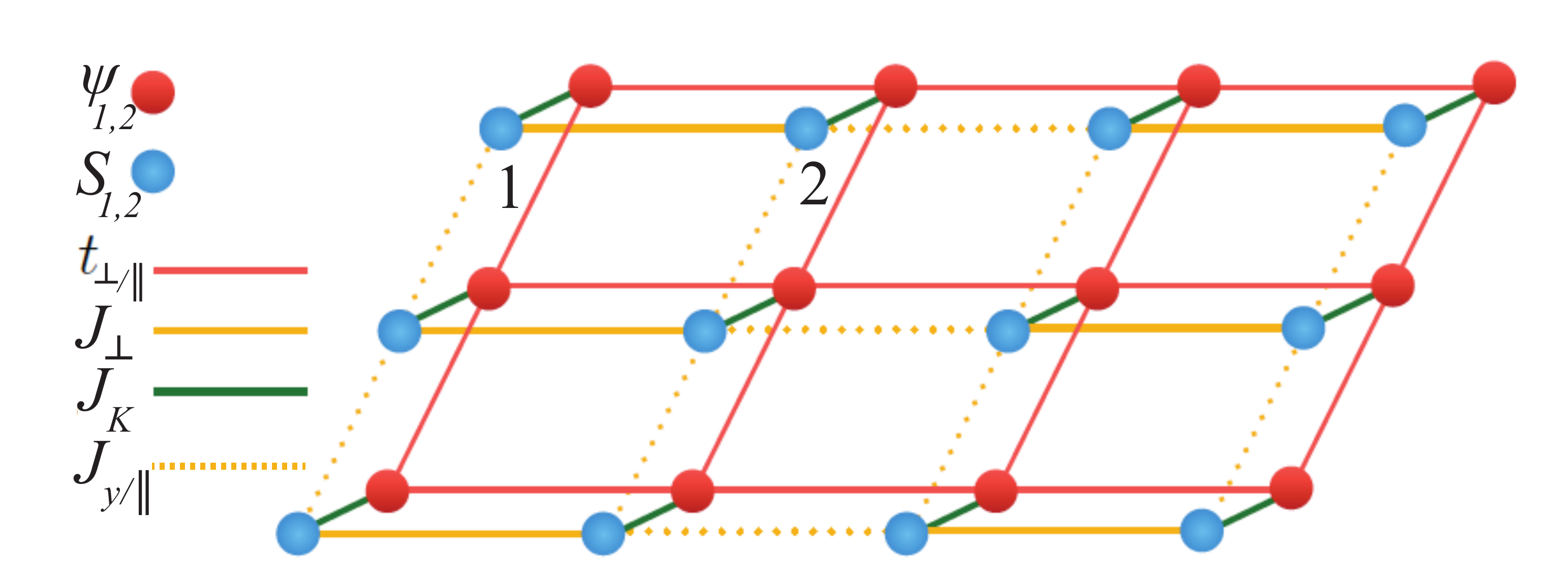}
	\caption{Structure of the 2D version of the model \ref{eq:hinit}.}
	\label{fig:2d}
\end{figure}
To discuss the Kondo coupling at low energies we need first to diagonalize the conduction electron hamiltonian. Denoting two sites of each rung by a sublattice index (1,2) we get the following hamiltonian:
\begin{equation}
-
\begin{bmatrix}
\psi_1\\
\psi_2
\end{bmatrix}^\dagger
\begin{bmatrix}
2t_\parallel \cos k_\parallel +2t_z \cos k^z  +\mu& t_\perp + t_y e^{2i k_y}\\
t_\perp + t_y e^{-2i k_y} & 2t_\parallel \cos k_\parallel +2t_z \cos k^z+\mu
\end{bmatrix}
\begin{bmatrix}
\psi_1\\
\psi_2
\end{bmatrix}.
\end{equation}
Note that we envision here a 3D lattice created by simply stacking the 2D layers in Fig. \ref{fig:2d} on top of each other in $z$ direction. Diagonalizing this we get:
\begin{gather*}
E_\pm = -2t_\parallel \cos k_\parallel -2t_z \cos k^z-\mu
\mp|t_\perp + t_y e^{2i k_y}|;
\\
\psi_\pm =
\frac{1}{\sqrt{2}}
\begin{bmatrix}
1\\
\pm(t_\perp + t_y e^{-2i k_y})/|t_\perp + t_y e^{2i k_y}|
\end{bmatrix}
\equiv
\frac{1}{\sqrt{2}}
\begin{bmatrix}
1\\
\pm e^{i\varphi_{k_y}}
\end{bmatrix},
\end{gather*}
where
\begin{equation}
e^{i\varphi_{k_y}} = \frac{t_\perp + t_y e^{-2i k_y}}{|t_\perp + t_y e^{2i k_y}|};
\;
\varphi_{k_y}|_{k_y\to \pi}\approx -2\frac{t_y (k_y-\pi)}{|t_\perp + t_y|}.
\label{eq:23dphase}
\end{equation}
One can reexpress the original $\psi_{1,2}$ fields as
$\psi_1({\bf k}) = [\psi_+({\bf k})+\psi_-({\bf k})]/\sqrt{2},\;\psi_2({\bf k}) = e^{-i\varphi_{k_y}}[\psi_+({\bf k})-\psi_-({\bf k})]/\sqrt{2}$. We now rewrite the Kondo coupling projected to the two low-lying states in magnetic field (\ref{eq:hkproj}):
\begin{gather*}
H_K \approx \frac{J_K}{2} \sum_{{\bf k},{\bf q}}(\tilde{S}^z({\bf q})+1/2) (s_{1}^z({\bf q})+s_{2}^z({\bf q}))
-\frac{J_K}{2\sqrt{2}}
\sum_{{\bf k},{\bf q}}
\tilde{S}^-({\bf q})(s_{1}^+({\bf q})-s_{2}^+({\bf q}))+\tilde{S}^+({\bf q})(s_{1}^-({\bf q})-s_{2}^-({\bf q}))
\\
s^\alpha_{a}({\bf q}) = \psi^\dagger_{a}({\bf k}+{\bf q})\sigma^\alpha \psi_{a}({\bf k}),
\end{gather*}
where introducing $\psi^\pm$ one gets:
\begin{equation}
\begin{gathered}
H_K =
\frac{J_K}{2} \sum_{{\bf k},{\bf q}}
(\tilde{S}^z({\bf q})+1/2)
\left(
\frac{1+e^{i(\varphi_{k_y+q_y} - \varphi_{k_y})}}{2}
[s^z_+({\bf q}) +s^z_-({\bf q})]+
\right.
\\
\left.
+
\frac{1-e^{i(\varphi_{k_y+q_y} - \varphi_{k_y})}}{2}
[\psi_+^\dagger({\bf k}+{\bf q})\sigma_z\psi_-({\bf k})+\psi_-^\dagger({\bf k}+{\bf q})\sigma_z\psi_+({\bf k})]
\right)
\\
-\frac{J_K}{2\sqrt{2}}
\sum_{{\bf k},{\bf q}}
\tilde{S}^-({\bf q})
\left(
\frac{1-e^{i(\varphi_{k_y+q_y}- \varphi_{k_y})}}{2}
[s^+_+({\bf q}) + s^+_-({\bf q})]+
\right.
\\
\left.
+
\frac{1+e^{i(\varphi_{k_y+q_y} - \varphi_{k_y})}}{2}
[\psi_+^\dagger({\bf k}+{\bf q})\sigma_+\psi_-({\bf k})+\psi_-^\dagger({\bf k}+{\bf q})\sigma_+\psi_+({\bf k})]
\right)
+h.c.
\end{gathered}
\label{eq:hk23d}
\end{equation}
From the definition \eqref{eq:23dphase} it follows that for $q_y\approx \pi$ $e^{i(\varphi_{k_y+q_y})}\approx e^{i(\varphi_{k_y})}+O(q_y-\pi)$; i.e. for $q_y\to \pi$ one finds that $\frac{1-e^{i(\varphi_{k_y+q_y} - \varphi_{k_y})}}{2}$ vanishes linearly as $q_y-\pi$, while $\frac{1+e^{i(\varphi_{k_y+q_y} - \varphi_{k_y})}}{2}\to1$.

We can study now the scaling dimensions of the Kondo coupling at a magnon BEC critical point where $(\tilde{S}^z({\bf q})+1/2) \to \int{dl}{2\pi} a^\dagger({\bf l}+{\bf q}) a({\bf l}), \tilde{S}^-({\bf q})\to a({\bf q})$. We assume that the BEC transition is caused by the condensation of bosons at the AFM wavevector $Q_0=(\pi,\pi)$ in 2D or $Q_0=(\pi,\pi,\pi)$ in 3D. We first remark that with bare dynamical exponent $z=2$ the ordinary bosonic quartic term is marginal in $d=2$ and irrelevant for $d>2$ (scaling dimension $d+z-4$) thus we can ignore corrections to the bosonic scaling in Eq. (\ref{eq:hk23d}).

We now proceed to the analysis of the Kondo coupling. As $z=1$ for the fermions, assigning scaling dimensions is not straightforward; we will thus consider several concrete cases. As the bosons condense at a finite wavevector $Q_0$ for a generic Fermi surface only points connected by $Q_0$ ('hot spots') take part in the low-energy physics at the QCP. Near a hot spot, the low-energy lagrangian of the fermions is $\int d\varepsilon d^d k\psi^\dagger ( -i \varepsilon + v_F k_1 + \sum_{i=2}^{d-1} a_i k_i^2 )\psi$, where we omit the spin/band indices in the Fermi fields for brevity.

We would like to analyze the relevance of Kondo coupling near the fixed point where $z=2$ bosonic behavior and the Fermi liquid physics of fermions are both intact. If the Fermi velocities at the 'hot spots' are collinear to each other then the momentum along the Fermi velocity should scale like energy ($[k_1] = [\varepsilon]$) to preserve the Fermi liquid behavior; on the other hand, to be consistent with the $z=2$ behavior one has to have $[k_1] = 2 [k_{i\neq1}]$ \cite{sachdev.1999}. Setting the scaling dimension of the momenta, perpendicular to the Fermi velocity to $1$, we get $[\psi] = -\frac{d-1+3z}{2} = -\frac{d+5}{2}$ and $[a] = - \frac{d-1+2z+2}{2} = -\frac{d+5}{2}$. The scaling dimension of a Yukawa-like coupling (corresponding the interband part of $H_K$ \eqref{eq:hk23d}) is then $\frac{3-d}{2}$. This result is also consistent with the patching scheme introduced in \cite{yamamoto.2010}; in that case, the scaling dimension of frequency/energy is set to $1$ resulting in an overall factor of $1/z$: $\frac{3-d}{4}$. Note that in the single-band case the interband term, as is shown above, vanishes linearly in momentum, and thus its scaling dimension is at least $\frac{3-d}{2} - 1 = \frac{1-d}{2}$. Lastly, a term of the form $\int \psi^\dagger\psi a^\dagger a$ arises from the ZZ part of the Kondo coupling; its scaling dimension is $1-d$. We have thus shown that the $z=2$ BEC QCP is stable in the one-band case for $d>1$ and there may be interaction effects at the 'hot spots' in the two-band case.

If the Fermi velocities at the 'hot spots' are not collinear to each other then the momentum along each Fermi velocity should scale like energy (as in the case of spin-density wave QCP \cite{sachdev.1999}), i.e.($[k_1] = [k_2] = [\varepsilon]$), while $[k_1] = 2 [k_{i>2}]$ to preserve $z=2$. Note that the latter is possible in $d>2$. Proceeding with this case we get the scaling dimensions $[\psi] = -\frac{d-2+4z}{2} = -\frac{d+6}{2} = [a]$. For Yukawa-like coupling we get $\frac{2-d}{2}$, while the intra-band term is always irrelevant. The $\int \psi^\dagger\psi a^\dagger a$ coupling has scaling dimension $-d$ and is also irrelevant. Thus in 3D the BEC QCP is stable in the one-band case. In 2D the situation is similar to spin-density wave transition \cite{sachdev.1999}, which also has $z=2$. The scaling scheme then requires $[\varepsilon]=2[k_1]=2[k_2]$, such that the bare frequency dependence of the fermions is dropped. The scaling dimension of a Yukawa-like coupling is then $\frac{2-d}{2}$; intraband coupling that vanishes linearly in momentum is irrelevant always and the quartic $\int \psi^\dagger\psi a^\dagger a$ term has scaling dimension $1-d$ and is irrelevant.

Let us summarize the above. For the critical behavior of the Kondo-coupled magnon BEC transition only the fermions near the Fermi surface points connected by the condensation wavevector $Q_0$ ('hot spots') are important. in the 1-band case the BEC transition is stable in 2D and 3D generalizations of the model; for the 2-band case interaction effects can be important in 2D and even in 3D for the case of collinear Fermi velocities at the hot spots.

\section{DMRG method and additional numerical results}

For numerical simulation, we employed the DMRG algorithm in its matrix product states (MPS) formulation using the ITensor package \cite{ITensor}. The presence of gapless modes identified with particle-hole excitations near the Fermi points and the softening magnon mode close to the BEC transition lead to a unfavorable logarithmic scaling of the Von Neumann entanglement entropy (EE) as a function of systems size. This in turn requires considering relatively large bond-dimensions and a careful monitoring of the convergences of different observables with increase in the bond-dimension. 

To showcase this procedure, in \cref{fig:xd_vs_chi}, we track the convergence of the intra-band super-conducting correlation, $\chi^+_D$ (see main text) as a function of the bond dimension. We present the two-band case, using the same microscopic parameters as the main text. A finite bond dimension restricts the capacity of the variational MPS leading to a spurious finite correlation length. We indeed find that progressively increasing the bond dimension $\chi$ provides a converged result that reproduces the expected power-law correlation on a finite size chain.

\begin{figure}
	\centering
	\includegraphics{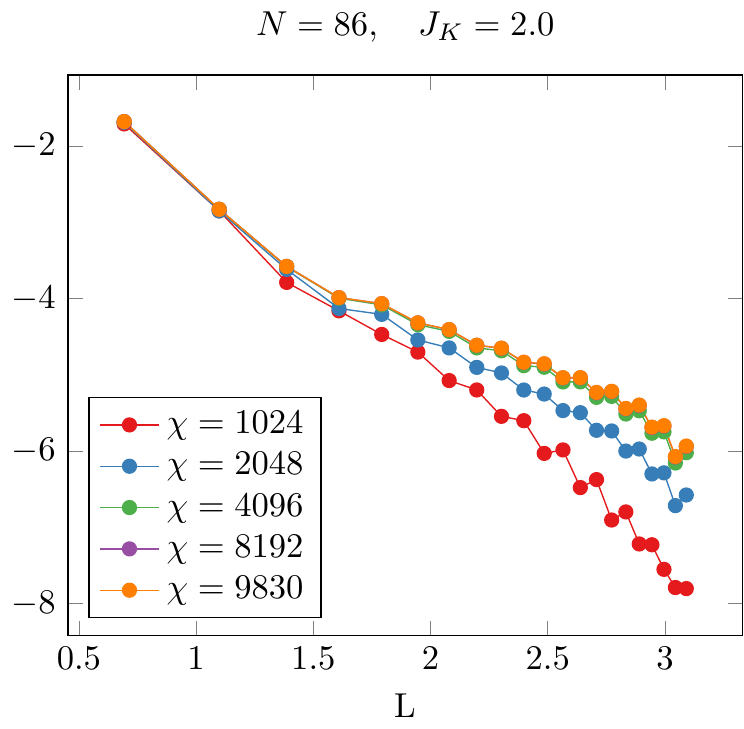}
	\caption{Convergence of $\chi^+_D$ as function of the bond dimension for $N=86$ and $J_K=2.0$, all other microscopic parameters are same as the ones used in the main text for the two-band case. }
	\label{fig:xd_vs_chi}
\end{figure}

To facilitate the computation, we explicitly conserve the $U(1)$ charges associate with the number of electrons and spin-rotations about the $z$ axis. In the presence of a finite Zeeman term, the ground state is not necessarily in the $M^z=0$ sector. Therefore, to determine the global ground state, we apply the DMRG algorithm on several low-lying  $M^z$ sectors  and identify the lowest ground-state energy. We note that since the Zeeman term does not couple to the total magnetization ($g_f\ne g_c$), one can not use a Legendre transformation to determine the ground state wave function \cite{Fouet_2006}.

{We begin by presenting our finite size scaling analysis used to determine the correlation length exponent, $\nu$. First, we estimate the finite-size value of the critical field $h_c(L)$, defined as the crossing position of $\Delta_s L^2$ curves corresponding to system sizes $L$ and $2L$. We then assume, the standard critical finite-size scaling anstaz, $\Delta_s L^2=R(\delta h L^{1/\nu})$, where $R(x)$ is a universal scaling function and $\delta h=h-h_c$. With these definitions, one can easily verify the expected scaling behavior of the derivative $R'(L)=\frac{\partial R}{\partial h}|_{h_c(L)}\sim a+b L^{1/\nu}$. In \cref{fig:fit_nu}, we carry out a numerical fit to the above form, which gives our numerical estimate $\nu=0.49(1)$.}

We now elaborate on our finite size-scaling analysis used to determine the spin gap in the thermodynamic limit. For non-interacting one dimensional fermions, the spin gap is expected to vanish linearly with inverse system size $\sim 1/L$. This fact follows directly from linearizing the the electronic dispersion near the Fermi points and taking in account the quantization of momentum in a finite size system. 

The above result is expected to hold asymptotically for large $L$. However, when studying the VBS metal phase on finite chains, we observed a sizable oscillatory behavior as a function of $L$, see left panel of \cref{fig:fit_1_N}. These oscillations hinder the extrapolation to the thermodynamic limit. To alleviate this issue, we consider a modified data set computed by a moving average over adjacent system sizes, marked by a blue curves in \cref{fig:fit_1_N}. This procedure significantly reduces the oscillations while keeping the fitting procedure well defined. We use the above outlined procedure to determine the energy level spacing $\Delta E(L)=E(M=1,L)-E(M=0,L)$ for $L\to \infty$.

\begin{figure}
	\centering
	\includegraphics[scale=1.0]{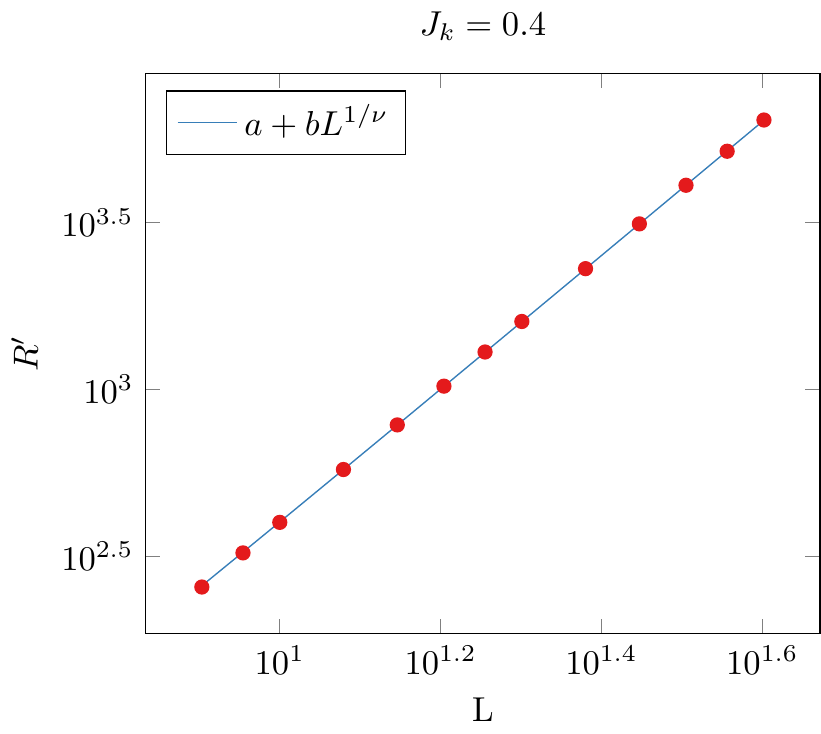}
	\caption{$R'$, see text for definition, as a function of $L$ shown on a log-log scale, for $J_k=0.4$ and all other parameters are the same as the single band case in the main text. Solid line is a fit to the finite-size scaling ansatz $a+b\times L^{1/\nu}$}
	\label{fig:fit_nu}
\end{figure}

\begin{figure}
	\centering
	\includegraphics[width=\linewidth]{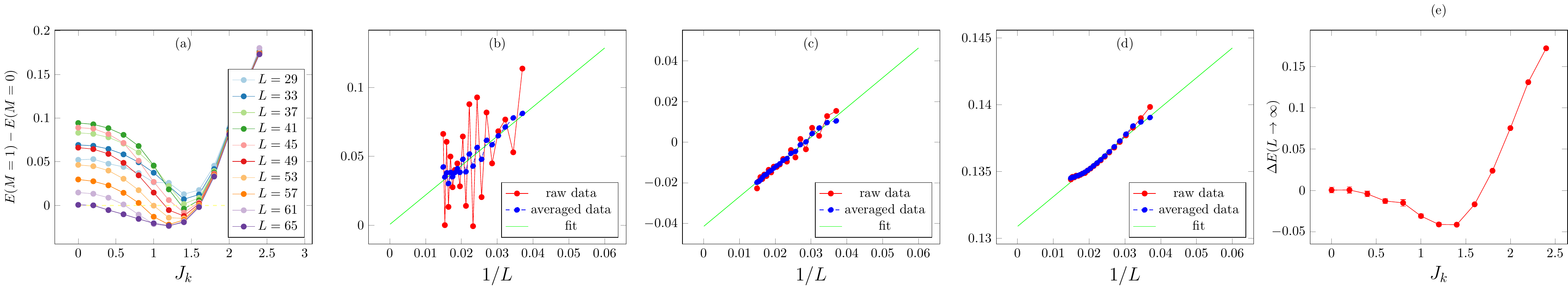}
	\caption{Finite size analysis of the energy difference $\Delta E(L)=E(M=1,L)-E(M=0,L)$. (a) $\Delta E(L)$ versus $J_K$, for different chain length $L$. (b)-(d) $\Delta E(L)$ as a function of inverse chain length for a fixed $J_K$. Red curves correspond to raw data, blue curves are a moving average, and green curves are a fit to the form $a+b/L$ (e) Extrapolated $\Delta E(L\to \infty)$ as a function of $J_K$.}
	\label{fig:fit_1_N}
\end{figure}

We note that for intermediate values of $J_K$, we can not rule out a magnetically polarized state since $\Delta E$ turns negative. However, the magnetization, inferred from our finite size data, is small and can be estimated to be no larger than $m\approx 1/100$ in absolute units.

\end{widetext}
\end{document}